\documentclass[letterpaper]{aa}
\usepackage{psfig}

\usepackage[dvips]{graphics}

\def\HII {H{\sc ii}}

\newcommand{\msol}{\mbox{$M_{\odot}$}}
\newcommand{\lsol}{\mbox{$L_{\odot}$}}

\newcommand{\kms}{\mbox{km~s$^{-1}$}}

\newcommand{\asec}{$^{\prime \prime} $}                   
\newcommand{\amin}{$^{\prime} $}
\newcommand{\mcu}[1]{\multicolumn{1}{c}{#1}}
\newcommand{\mcd}[1]{\multicolumn{2}{c}{#1}}

\begin{document}
\title{New light on the S235A-B star forming  region\thanks{Based on 
observations carried out with the IRAM Plateau de
Bure Interferometer. IRAM is supported by INSU/CNRS (France), MPG
(Germany) and IGN (Spain)}
}

\author{Marcello Felli\inst{1}, Fabrizio Massi\inst{1}, Alessandro 
Navarrini\inst{2,3},  Roberto Neri\inst{3}, Riccardo Cesaroni\inst{1}, 
Tim Jenness\inst{4} 
}
\offprints{M. Felli, mfelli@arcetri.astro.it}

\institute{ INAF-Osservatorio Astrofisico di Arcetri, 
Largo E. Fermi, 5, 50125 Firenze, Italy
\and
Radio Astronomy Laboratory, University of California, Berkeley 601 Campbell
Hall, Berkeley, CA 94720, USA
\and
IRAM, 300 Rue de la Piscine, Domaine Universitaire, F-38406 St. Martin
d'H\`eres Cedex, France
\and
Joint Astronomy Centre, 660 N.  A'oh\={o}k\={u} Place, Hilo, HI 96720, USA
}

\date{Received; Accepted }

\abstract{
The S235A-B star forming region has been extensively observed
in the past from the radio to the near IR, but what was happening in the
immediate surroundings of the water maser, placed in between the two
nebulosities,  was still unclear because
of insufficient resolution especially in the spectral range 
from the Far IR to the mm, even though
there were sound indications that   new young stellar objects  (YSOs)
are being formed there. 
We present here new high resolution maps at mm wavelengths  in different
molecules (HCO$^+$, C$^{34}$S,
H$_2$CS, SO$_2$ and CH$_3$CN), as well as in the 1.2 and 3.3 mm
continuum obtained with the Plateau de Bure interferometer, and 
JCMT observations at 450 ${\mu}$m and 850 ${\mu}$m that 
unambiguously reveal the presence of new  YSOs placed in between
the two \HII\ regions S235A and S235B and associated with the water maser.
A molecular core and an unresolved source in the  mm and in the sub-mm 
are centred on the maser, with indication of mass infall onto the core. 
Two molecular bipolar outflows  and a
jet originate from the same position. Weak evidence is found for 
a molecular rotating disk 
perpendicular to the direction of the main bipolar outflow. 
The derived parameters  indicate that one of the YSOs is an intermediate
luminosity object ($L \sim10^3L_{\odot}$) in a very early evolutionary phase,
embedded in a molecular core of $\sim100~M_{\odot}$, with  a temperature of 
30 K. The main  source of energy for the YSO  could come  from
gravitational infall, thus making of this YSO  a rare example of intermediate
luminosity protostar representing a link between  
the earliest evolutionary phases of massive stars
and low mass protostars of class 0--I.
}

\authorrunning{M. Felli et al.}
\maketitle

\section{Introduction}

S235A and S235B are two close-by (separation $\sim$1\amin) 
small patches of nebulosity ($\sim$20\asec\ and $\sim$10\asec\ in diameter, 
respectively), so called because they are located about 10\amin\ south 
of the more extended \HII\ region S235 (Sharpless~\cite{S}; Felli \& 
Churchwell~\cite{FC}), but most 
probably not directly related to it (Israel \& Felli~\cite{IF}).
About 3.5 arcmin south of S235A-B there is another patch of nebulosity
called S235C and identified with a small \HII\ 
region (Israel \& Felli~\cite{IF}).

Early interferometric radio continuum observations have shown that 
S235A is a young \HII\ region, even though far beyond the earliest
ultracompact (UC) evolutionary phase, while no radio continuum emission
comes from S235B (Israel \& Felli~\cite{IF}).  
The interest in these two nebulosities was increased by the fact
that an H$_2$O maser is present in the same area (Henkel et 
al.~\cite{HHG}; Comoretto et al.~\cite{COMORE}), a well known  landmark
of recent star formation (see e.g. Tofani et al.~\cite{tof95}). With
the resolution available at that time it was not clear whether  the maser
was associated with S235A, S235B or simply happened to be
in the same  star forming region (Israel \& 
Felli~\cite{IF}). 
Single dish observations have also revealed the presence of 
a methanol maser (Nakano \& Yoshida~\cite{NY}; Haschick et al.~\cite{HMB})
and of an SiO maser (Harju et al.~\cite{HLBZ}).

To ascertain the true relationship between S235A-B and the water 
maser and establish
which is the source of energy supply for the water maser emission, 
high resolution radio observations  were performed 
with the VLA in the A configuration in the water line and in the continuum
(Tofani et al.~\cite{tof95}), 
which led to  the result  that the  H$_2$O maser is well
outside the boundary of the S235A \HII\ region, away from S235B
and that no point-like radio
continuum source (down to a flux density of $\sim$0.5 mJy and 
with a resolution
of 0\farcs3) is 
associated with  it, i.e. no trace
of a UC \HII\ region near the maser was found.

At that time it was  becoming clear that  water masers without 
associated compact radio continuum emission could trace one of the earliest 
phases in the formation of a young stellar object (YSO), in which 
the UC \HII\ region ionized by the early type star  (if at all present)
is so dense and optically thick
to escape detection at radio wavelengths (Codella et al.~\cite{cfn},
\cite{ctc}).

The presence of a molecular cloud in the S235A-B region
was known from earlier low resolution observations in several molecular lines
(Evans \&  Blair~\cite{EB}; Lafon et al.~\cite{LDBL};
Nakano \& Yoshida~\cite{NY}).
To test the hypothesis that the water maser indicated the
location of one or more new  YSOs independent from S235A-B, 
and get more information on the environment
in which the water maser was located, new maps in molecular tracers
of high density gas (Cesaroni et al.~\cite{CFW}) and
near-IR observations of the region (Felli et al.~\cite{FTVW}) were made.

\begin{figure}
\centerline{\psfig{figure=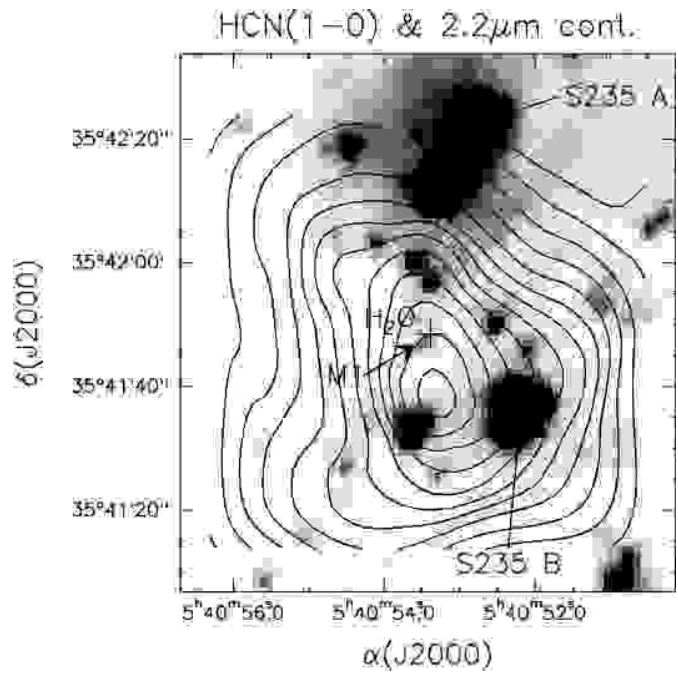,width=8.8cm}}
\caption[]{Overlay of the Pico Veleta HCN map (contours, resolution 27\asec)  
from Cesaroni et al.~(\cite{CFW}) with the TIRGO K-band 
image of the S235A-B region (grey scale)
from Felli et al.~(\cite{FTVW}). The VLA position
of the water maser from Tofani et al.~(\cite{tof95}) is indicated with a 
cross (much larger than the positional uncertainty which is 0.1\asec).
M1, the K-band  source with the largest infrared excess, is also indicated.
}
\label{hcnk}
\end{figure}

The results confirmed the validity of the hypothesis: 
the water maser is located at the centre
of a high density small diameter molecular core, 
which peaks just between
the S235A and S235B nebulosities. A cluster of near-IR sources  is  also
located in the same area. The source with the largest IR excess
(called M1)
is  placed close to the water maser. The picture before the
present observations is summarized in Fig.~\ref{hcnk} which shows the
overlay of the HCN map obtained with  a resolution of 27\asec\
(Cesaroni et al.~\cite{CFW}) with a K-band (2.2 ${\mu}$m) 
map of the same area (Felli et al.~\cite{FTVW}).
All the indications suggest  that the water maser,  M1 (both indicated in
Fig.~\ref{hcnk}) and the HCN clump
are all part of a much younger episode of star formation 
with respect to the S235A-B nebulosities, not necessarily directly 
related to them. 

For completeness it should be noted that the VLA observations of the water
maser (Tofani et al.~\cite{tof95})
refer to the component at $ \sim-60$ \kms, which happened to be
flaring at the time of the  observations (1992) and was the only feature
above the noise in the observed bandwidth (from --78 to --40~\kms). 
During the continuing patrol of the
water maser emission with the Medicina radiotelescope, started in 1988,
this velocity component has very seldom emerged again from the noise level,
while two other velocity components between  $ \sim-2.5$  and 2.5 \kms\ 
are frequently observed (Felli et al., in preparation). A partial report of the 
results of the Medicina patrol is  given in Felli et al.~(\cite{FTVW}).

In the present work we 
investigate with better resolution the region around the water maser
to study in more
detail the YSOs that presumably are  harboured in the centre of the
molecular cloud.
We present arcsec resolution maps at mm wavelengths  in different
molecules that trace high density gas (HCO$^+$, C$^{34}$S,
H$_2$CS, SO$_2$ and CH$_3$CN), as well as in the 1.2 and 3.3 mm
continuum sensitive to  thermal dust emission.

To obtain the bolometric luminosities of the YSOs and their spectral energy
distribution (SED), high resolution far IR observations are
needed, since at these wavelengths YSO spectra are  expected to have 
their maximum. We inspected the IRAS HIRES  maps (Aumann et al.~\cite{afm}) 
of the region around  the water
maser (which contains IRAS\,05375+3540) 
in the 
four IRAS bands, but the resolution is  inadequate for separating the contribution
of S235A and  S235B from that of the YSO. 

We also present here archive JCMT
observations at 450 ${\mu}$m and 850 ${\mu}$m that 
unambiguously demonstrate that the sub-mm emission peaks on the YSO 
associated with the water maser
and separate its emission  from that of S235A-B.

For consistency with previous work a distance of 1.8 kpc will be assumed 
(Nakano \& Yoshida~\cite{NY}).

%
%

\section{Observations and results}
\label{obs}

\subsection{Plateau de Bure observations}
\label{pdb}

The interferometric observations were carried out on November 22, 1997 and 
March 25, 1998 in the CD set of configurations of the IRAM five 
element array at Plateau de Bure, France (Guilloteau
et al.~\cite{iram}).
The five 15-m antennas were equipped with 82--116~GHz and 210--245~GHz
SIS receivers operating simultaneously with double-sideband (DSB) 
noise temperatures of 35 K at 3.3 mm and 50 K at  1.2 mm. The receivers were
tuned DSB at both wavelengths. The facility correlator was centred at 
89.014  GHz in the lower side-band  at 3.3 mm 
and at 241.102 GHz in the lower side-band
at 1.2 mm.
We show the frequency set-up of the correlator and list the 
detected spectral lines in Table~\ref{tfreq}.
The six units in the correlator were placed in such a way that a frequency 
range free of lines could be used to measure the continuum flux. 
The effective spectral resolution is about twice the channel spacings
given in Table~\ref{tfreq}. Source visibilities were phase calibrated
by means of interspersed observations (every 20 minutes) of 
nearby point sources.

\begin{table*}
\begin{flushleft}
\caption[]{Frequency set-up for the Plateau de 
Bure interferometer.
 }
\label{tfreq}
\begin{tabular}{lcrll} 
\hline
Line & \mcu{Frequency}    & \mcu{Band} & \mcd{Channel spacing$^a$} \\
     & \mcu{(MHz)} & \mcu{(MHz)} & \mcu{(MHz)} & \mcu{(\kms)} \\
\hline
3.3~mm continuum       & 90494 &  160~  & ~2.5      & ~8.29  \\ 
HCO$^+$(1--0)           & 89189 &  20~  & ~0.078125 & ~0.26 \\ 
CH$_3$CN(5--4)       & 91987 &  40~  & ~0.15625    & ~0.51  \\ 
1.2~mm continuum$^b$   & 242631 &  160~  & ~2.5  & ~3.09  \\ 
C$^{34}$S(5--4)           & 241016 & 40~  & ~0.15625      & ~0.19   \\ 
H$_2$CS 7(1-6)--6(1,5)  & 244048 & 160~  & ~2.5      & ~3.07    \\ 
SO$_2$ 14(0,14)--13(1,13) & 244254 &  40~  & ~0.15625    & ~0.192   \\ 
\hline
\end{tabular}

\vspace*{1mm}
$^a$~the effective spectral resolution is about twice the nominal channel
 spacing \\
$^b$~the 1.2~mm continuum was obtained by averaging two 160~MHz
 bandwidths in the lower and upper side bands
\end{flushleft}
\end{table*}

The bandpass, amplitude and phase calibration was carried out in the 
standard antenna based manner.
The flux of the primary calibrator was bootstrapped from IRAM
monitoring measurements and used to derive the absolute flux density scale.
Table~\ref{tinstr}  gives a list
of the main parameters for our Plateau de Bure interferometer  observations.
The phase centre position is that of the water 
maser (Tofani el al.~\cite{tof95}).

\begin{table*}
\caption[]{Instrumental parameters for the IRAM Plateau de Bure interferometer
observations}
\label{tinstr}
\begin{flushleft}
\begin{tabular}{lc}
\hline
\mcu{Parameter} & \mcu{Value} \\
\hline
Centre of phase (water maser) & \mcu{$\alpha(2000) = 05^{\rm h} 40^{\rm m} 53\fs420$} \\
                & \mcu{$\delta(2000) = 35\degr 41\arcmin 48\farcs80$} \\
Observing mode & continuum+line; double  side-band at 3.3~mm and at
 1.2~mm\\
Number of antennas & 5 \\
Baseline range & 15--150~m \\
Band centre & 89.014~GHz (LSB); 241.102~GHz (LSB)\\
Number of sections in the correlator & 6 \\
Primary HPBW  & 55\asec; 20\asec \\
Synthesised HPBW (P.A.) & 3.3~mm: 5\farcs4 $\times$ 4\farcs2~(61\degr) \\
                        & 1.2~mm: 2\farcs2 $\times$ 1\farcs8~(45\degr) \\
Primary flux density and & 0415+379 (5.5--6.7~Jy at 3.3~mm;
                                         3.4--4.1~Jy at 1.2~mm) \\
bandpass calibrators     & 3C\,273 (25.3--29.8~Jy at 3.3~mm;
                                        13.2--26.2~Jy at 1.2~mm) \\
                         & 3C\,345 (4.1~Jy at 3.3~mm; 1.1~Jy at 1.2~mm) \\
Phase calibrators & 2013+370 (1.8--2.2~Jy at 3.3~mm; 0.4--1.3~Jy at 1.2~mm) \\
                  & 2005+403 (0.65--0.75~Jy at 3.3~mm; 0.08--0.24~Jy at 1.2~mm)
\\
\hline
\end{tabular}
\end{flushleft}
\end{table*}

The calibration and data reduction were made using the
GAG software developed at IRAM and Observatoire de Grenoble.
Continuum subtraction was performed in the UV plane by using the integral
over the line-free channels of the 160~MHz units. Finally, channel maps
were produced for all the lines.
The conversion factor from flux to brightness temperature in the
synthesized beam is $\sim5.9$~K~(Jy/beam)$^{-1}$ at 3.3~mm and
$\sim5.8$~K~(Jy/beam)$^{-1}$ at 1.2~mm.

\subsubsection{Continuum}
\label{mm}

\begin{figure*}
\centerline{\psfig{figure=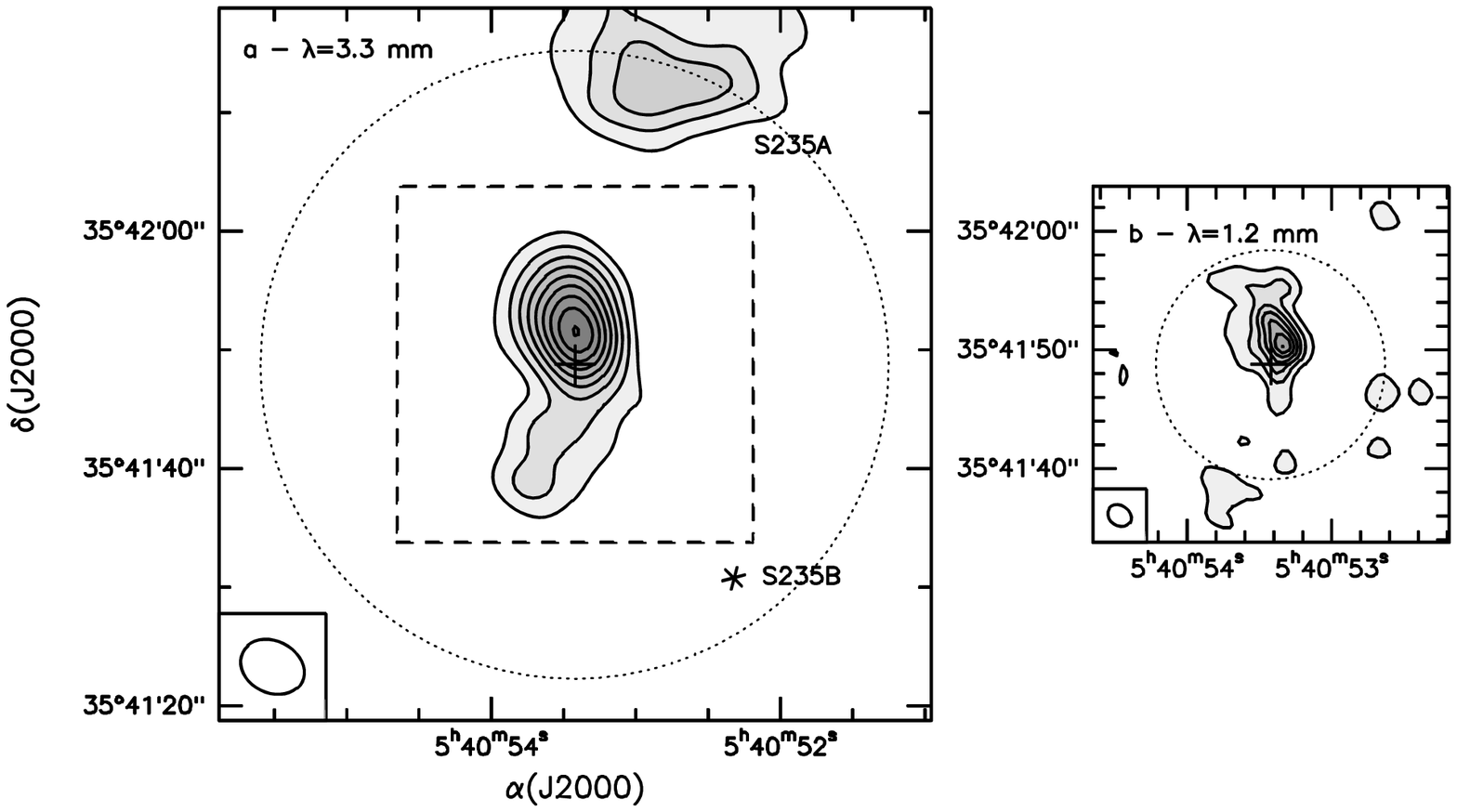,width=16cm,angle=-90}}
\caption[]{ The 3.3 mm (a) and 1.2 mm (b) maps. 
The water maser position is indicated 
by a cross. The dotted circle defines the primary beam HPW. 
The box in the lower left shows the synthesized HPBW.
The position of S235A (detected) and S235B (undetected) are marked in the 
3.3 mm map. Contour levels range from 1 (3 $\sigma$) to 9 by 1 
mJy/beam at 3.3 mm
and from 3 (3 $\sigma$) to 57  by 9 mJy/beam at 1.2 mm. 
The dashed square box in the 3.3 
mm map delimits the same area displayed in the 1.2 mm map.
}
\label{mmmap}
\end{figure*}

The 3.3 mm and 1.2 mm continuum maps (uncorrected for primary beam attenuation)
are shown in  Fig.~\ref{mmmap}. 

In the 3.3 mm map, S235A is clearly detected as an extended source 
at the edge of the
primary beam.
The total flux density (corrected for primary beam attenuation) is
54 mJy. This value is lower than the optically thin
extrapolation of the free-free emission observed at lower
frequencies ($\sim$200 mJy using the flux densities of
Israel \& Felli~\cite{IF}), but  confirms that the radio continuum 
emission comes from the ionized
gas of the S235A \HII\ region, with negligible  contribution from dust emission. 
The reason for the lower observed flux density must probably be
searched for in the higher resolution 
of the present observations (5$\farcs6$
compared with 12\asec) which
misses lower surface brightness extended emission. 
The overlay with the K-band image (Felli et al.~\cite{FTVW}), not
shown here, indicates that the S235A  mm emission is centred on the
western edge of a  K-band nebulosity, where a sharp drop in 
surface brightness suggests the presence of a foreground  dust front.

S235B is not detected at 3.3 mm
with an upper limit to a point source
at the same position of 1 mJy and  also undetected at 1.2 mm, but in this
case it is also well outside the primary beam.  
The nondetection  is consistent with previous 
upper limits at lower frequencies (Israel \& Felli~\cite{IF}) and confirms
that no radio emission  either from ionized gas or dust  is coming from S235B.
This behaviour of S235B at radio wavelengths is surprising, since the
source has been detected in H${\alpha}$ (Glushkov et al.~\cite{GDK})
and  Br${\gamma}$  (Krassner et al.~\cite{KPSH}).  
Possible explanations
for this effect in terms of an ionization-bounded
expanding envelope have already been discussed  by Felli et al.~(\cite{FTVW}).

S235C is totally outside the primary beam both at 3.3 and 1.2 mm.

%
%

The new important result of the present observations, in connection with
the hypothesis of YSOs associated with the water maser, is the detection
at 3.3 and 1.2 mm of an almost unresolved core slightly to
the north (2\asec) 
of the water maser. At the highest resolution of the 1.2 mm observations
the core diameter is 3\asec, only slightly larger than the synthesized
beam.  There is also  a tail protruding approximately 
to the south  of the core, more evident in the 3.3 mm map, but also
partly visible in the 1.2 mm map, extending for $\sim$10\asec\ in the
longest direction and almost unresolved in the other.

Because of the core-tail
morphology we have tried to separate the flux density of  
the core from that of the tail. 
The flux densities of the core (corrected for primary beam attenuation)
are 20  and 245~mJy at 3.3 and 1.2 mm, respectively.
The spectral index between 3.3 and 1.2 mm 
is 2.5. Since we cannot estimate the possible contribution
to the 3.3 mm continuum by free-free emission,
this is a lower limit to the dust spectral index,  but it 
clearly indicates
that most of the emission comes from the dust that surrounds the YSOs.
The extrapolated flux density of the core dust emission at 8.4 GHz is
0.06 mJy, well below the sensitivity of the  VLA  observations of Tofani
et al.~(\cite{tof95}). Thus the contribution of dust emission at 8.4~GHz
is likely to be negligible with respect to free-free emission, so that the
upper limit of 0.5 mJy (Tofani et al.~\cite{tof95}) can still be used to
constrain the emission of a UC \HII\ region at the position of the water
maser.

For the tail, the flux densities are 7.0  and 13  mJy at 3.3 and 1.2
mm, respectively, with some  uncertainty in the 1.2 mm flux 
density because of the weakness of the emission. The spectral
index of the tail between 3.3 and 1.2 mm is 0.6, definitely different
from that of the core and identical to that expected 
either from an ionized  stellar wind (Panagia \& Felli~\cite{PF})
or a thermal jet (Reynolds~\cite{R}),  and suggests that the
emission comes from an ionized jet, presumably departing from the
core, with very small  or null contribution from  dust emission.
%
As a counter-check, the  predicted flux density of the  jet at 8.4 GHz
using a spectral index 0.6 
is of the order of 1.7 mJy, so the fact that it was not detected 
in previous observations at 
lower frequencies is consistent with their sensitivities and
resolution. Deeper radio continuum observations at cm wavelengths  
are essential to better define the parameters  of the thermal jet.

\subsubsection{HCO$^+$}
\label{hco+}

\begin{figure}
\centerline{\psfig{figure=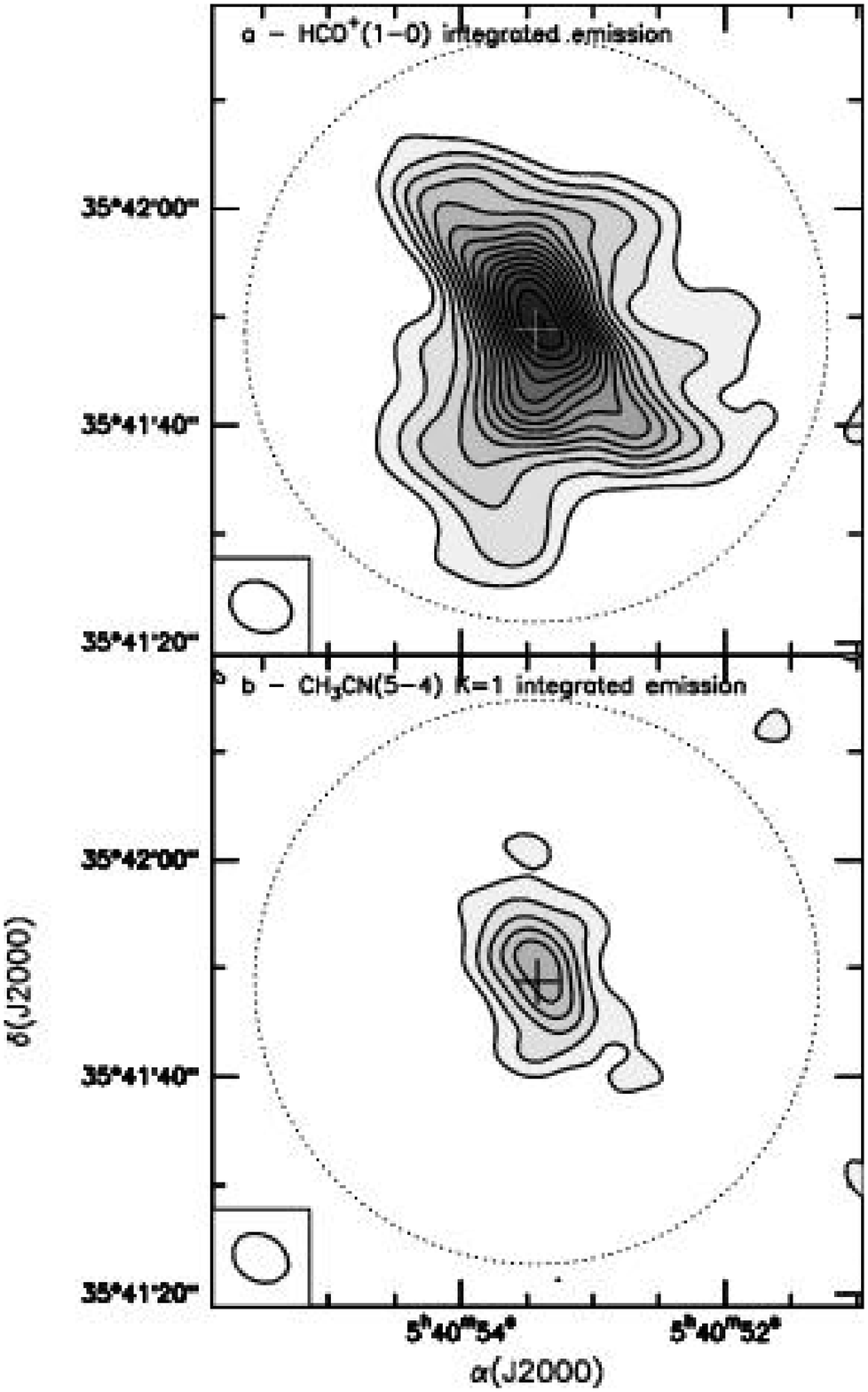,width=8.8cm}}
\caption[]{{\rm a.} The HCO$^+$ map averaged over the entire velocity range. 
The water maser position is indicated
by a cross. The scale is identical to that of the 3.3 mm map 
in Fig.~\ref{mmmap}. The dotted circle defines the primary beam HPW.
The box in the lower left shows the synthesized HPBW.
Contour levels range from 10 to 150 by 10 mJy/beam. 
{\rm b.} The CH$_3$CN(5--4)  map averaged over the velocity range
of the $K$=1 transition.
Contour levels range from 6 to 30 by 6 mJy/beam.
}
\label{hcoint}
\end{figure}

The HCO$^+$(1--0) map of the emission
averaged over the entire velocity range (from
--49 to 11.5 \kms) is shown in Fig.~\ref{hcoint}a.
The line emission peaks at the position of the water maser, making
tighter  the connection between water maser and molecular core  than 
what was found at  lower resolution in other molecular lines  
(Cesaroni et al.~\cite{CFW}).  The molecular 
core has a FWHM diameter of $\sim$ 11\asec\ 
and is surrounded by a more extended lower 
intensity region with a FWHM size  of $\sim$ 30\asec.

Comparison of the HCO$^+$ and the 3.3 mm continuum maps 
does not show any obvious evidence of  connection 
between the molecular core and
S235A (e.g.  no sharp
rim either in the molecular cloud or in the \HII\ region at the location where
the two are closest), thus making it very unlikely that the
ionized gas of S235A and the HCO$^+$ molecular cloud might
be interacting.

%
%

\begin{figure}
\centerline{\psfig{figure=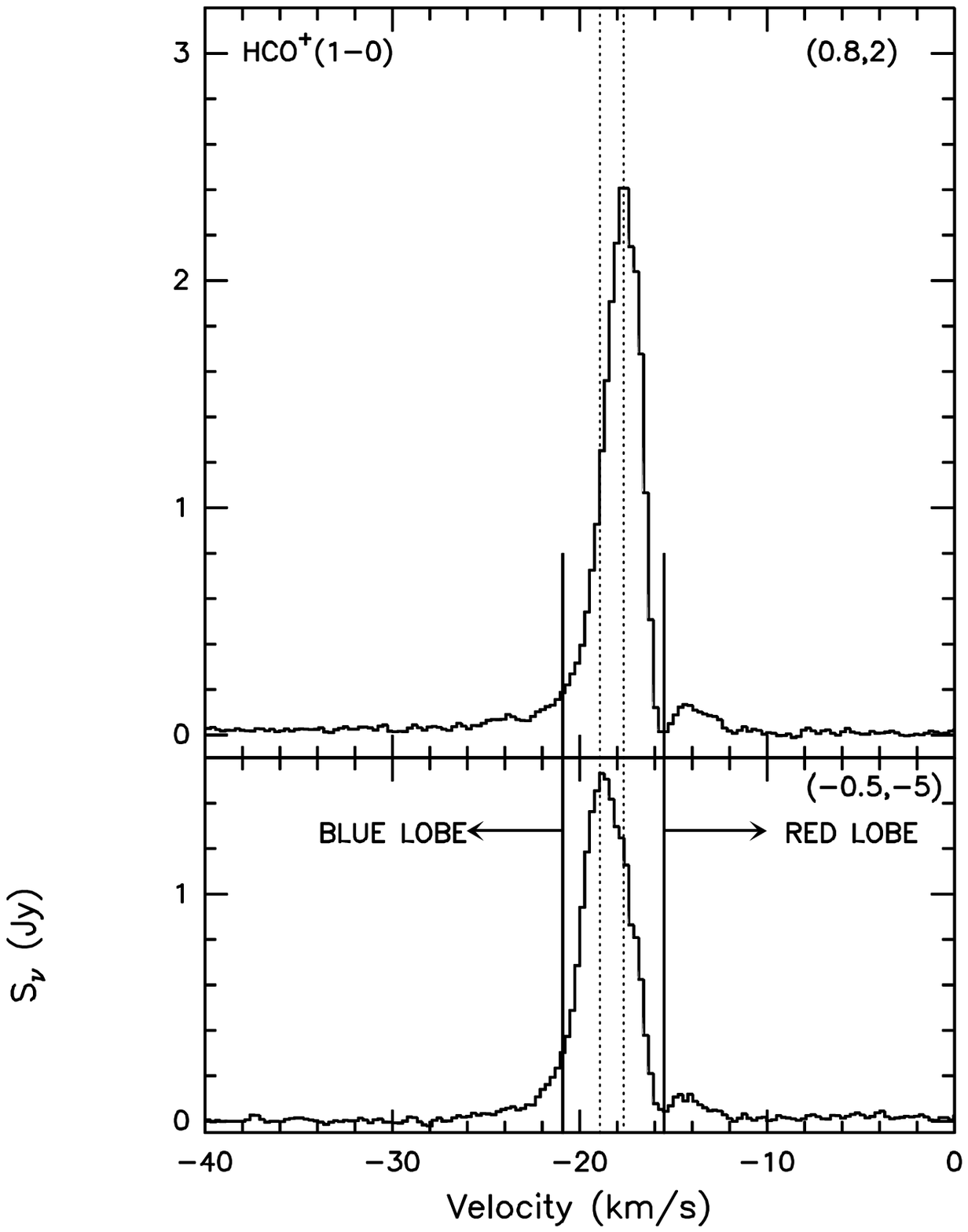,width=8.8cm}}
\caption[]{The HCO$^+$ line  profiles, averaged
over an area roughly equal to that of the synthesized beam, 
in two positions close to the peak.
The offsets (in arcsec) with respect to the water maser position are given
in the top right corner of each panel. The two dotted lines indicate the peak 
velocity at the two positions. The  two full lines indicate 
the limiting velocities 
used to define the blue and red outflow lobes.
}
\label{profiles}
\end{figure}

Inspection of the line profiles at two different positions
near the water maser  (see  Fig.~\ref{profiles}) 
reveals three  distinct
features: 1) extended wings  suggesting the presence of an outflow;
2) a shift in the velocity of the  peak with position, which 
suggests the presence of two distinct molecular components  
at  different velocities; and 3) a  dip  of emission 
on the red side of the profile  at --15.5  \kms, suggestive of red-shifted
self-absorption and hence of infall inside the core.

%
%

\begin{figure}
\centerline{\psfig{figure=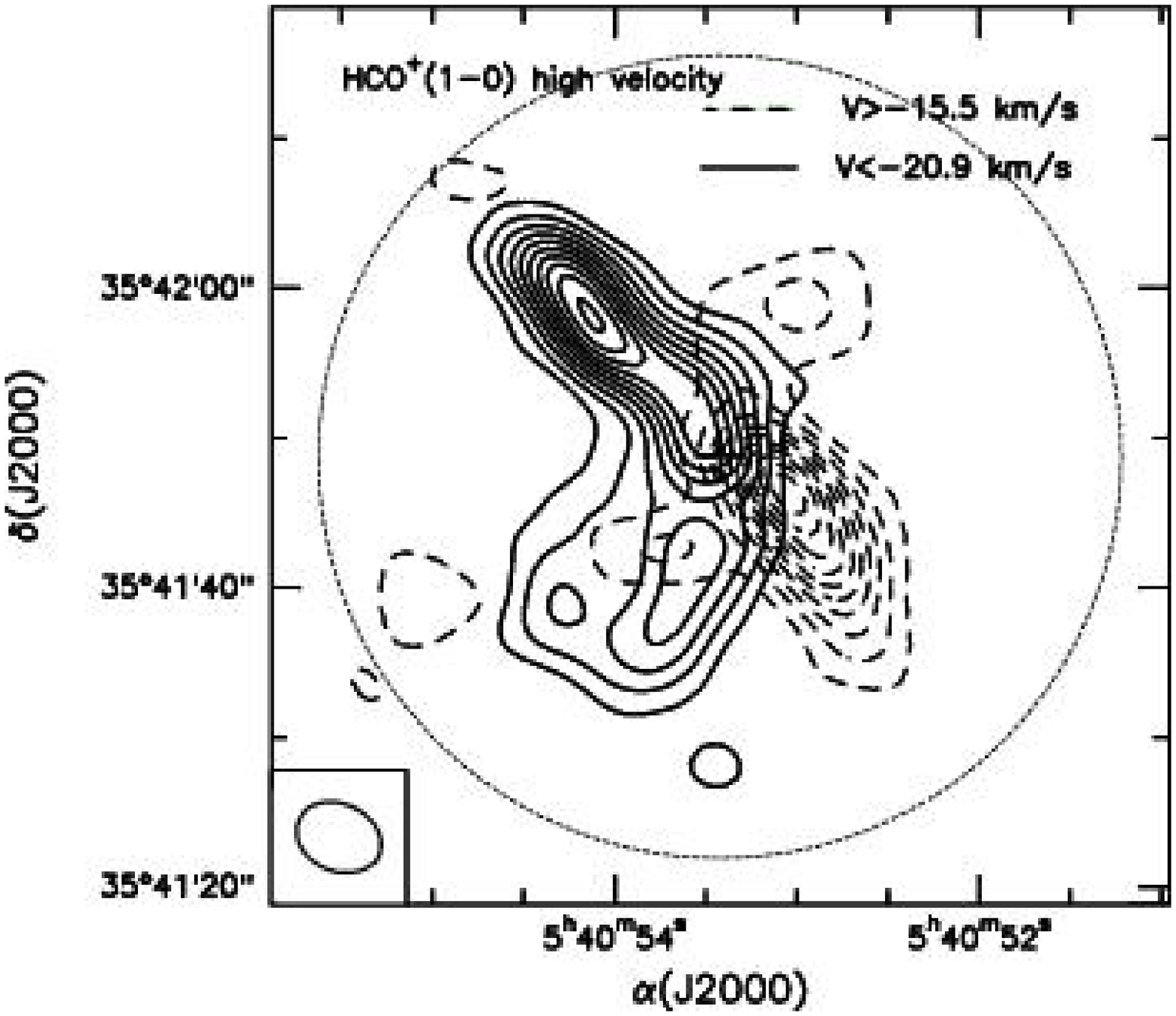,width=8.8cm}}
\caption[]{ The HCO$^+$ outflow.  
The water maser position is indicated
by a cross. The scale is identical to that of the 3.3 mm map
in Fig.~\ref{mmmap}. The dotted circle defines the primary beam HPW.
The box in the lower left shows the synthesized HPBW.
Contour levels range from 10 to 100 by 10 mJy/beam (red lobe, dashed)
and 6 to 60 by 6 mJy/beam (blue lobe, full line).
}
\label{outflow}
\end{figure}

To study the morphology of the outflow we have inspected
carefully the channel maps and produced a map
averaging all channels up to the edge of the band  
with  $v_{\rm LSR} >$ --15.8 \kms\ for the red lobe 
and with  $v_{\rm LSR} <$ --20.8 \kms\ for the blue lobe (see full lines
in Fig.~\ref{profiles}). The result is 
shown in Fig.~\ref{outflow}. Two bipolar outflows are evident:
1) a more extended one oriented in the NE--SW direction and 2)
a smaller one oriented in the NNW--SSE direction.
Within the uncertainty on the separation of the two outflows, 
both seem to originate from the position of the water maser.
The flux contained in the NE--SW  outflow lobes is $\sim$12\% of that
over the entire velocity range.
The presence of a larger scale bipolar outflow in the S235A-B region
had also been reported by Nakano \& Yoshida~(\cite{NY}) using the 
$^{12}$CO line and by Cesaroni et al.~(\cite{CFW}) and Felli et 
al.~(\cite{FTVW}) with $^{13}$CO observations. 
However there is very little correspondence between those maps and the map of
Fig.~\ref{outflow}: most probably the different resolutions and the fact that
the CO line is much more affected by the large scale, lower density molecular
gas make it impossible to compare the results of Cesaroni et al.~(\cite{CFW})
and Felli et al.~(\cite{FTVW}) to the map in Fig.~\ref{outflow}.

%
%

\begin{figure*}
\centerline{\psfig{figure=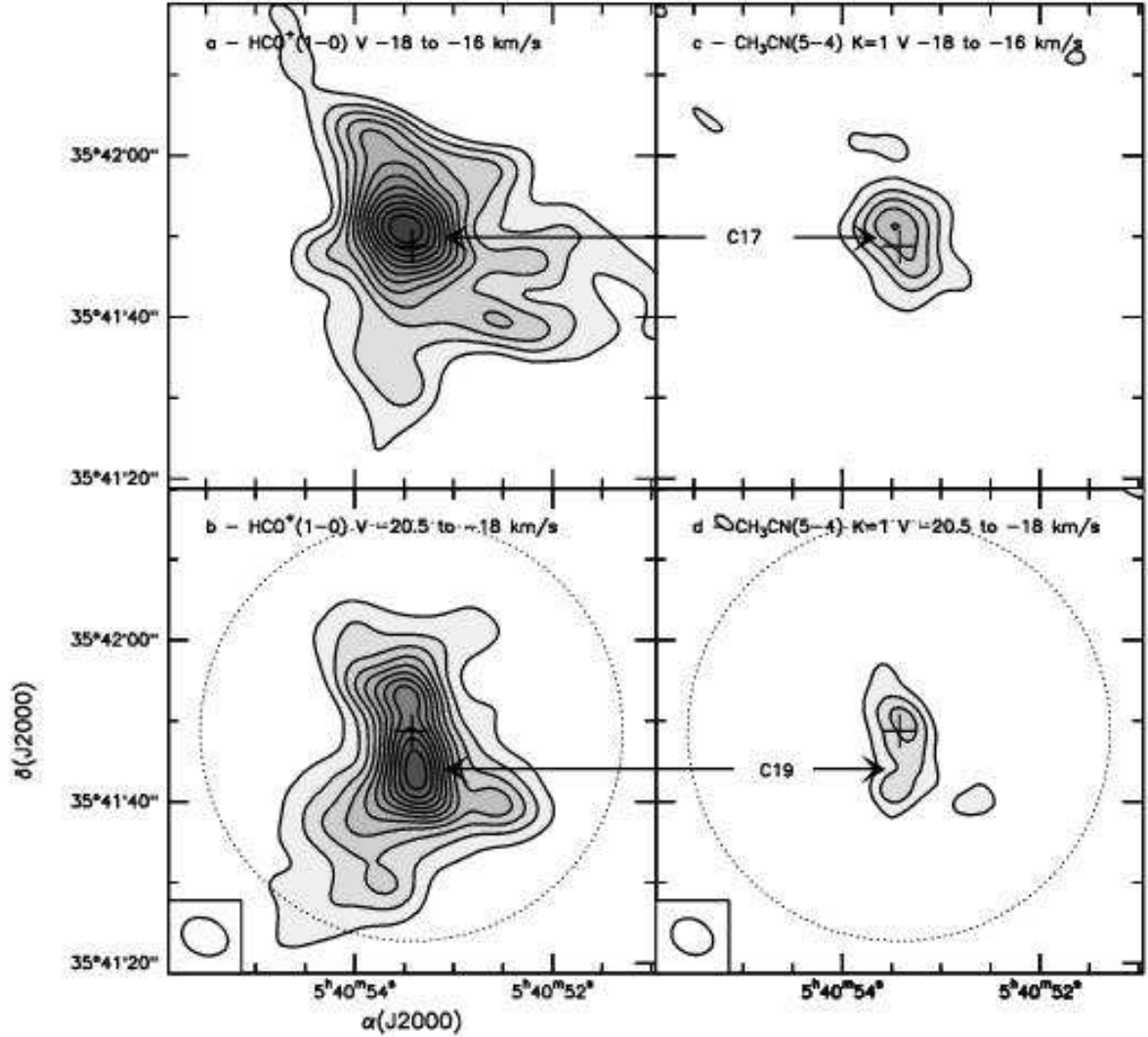,width=16cm,angle=-90}}
\caption[]{Maps of the line emission averaged in different velocity ranges. 
The water maser position is indicated
by a cross. The scale is identical to that of the 3.3 mm map
in Fig.~\ref{mmmap}. The dotted circle defines the primary beam HPW.
Left panel: HCO$^+$ emission: {\bf a.} C17; {\bf b.} C19.
Contour levels range from 0.15 to 1.8 by 0.15 Jy/beam (C17)
and 0.06 to 1.26 by 0.12 Jy/beam (C19).
Right panel:  CH$_3$CN emission: 
{\bf c.} C17; {\bf d.} C19.
Contour levels range from 0.01 to 0.05 by 0.01 Jy/beam.
}
\label{peaks}
\end{figure*}

To investigate the origin of the emission at different velocities (see
Fig.~\ref{profiles}),
we produced two maps in two  small velocity intervals centred around 
the  velocities marked with dotted lines in Fig.~\ref{profiles},
(approximately  --17 and --19 \kms).
The two maps are shown in Fig.~\ref{peaks}a--b and cover the velocity
intervals from --18.16 to  --15.8 \kms\ and from --20.8 to --18.16 \kms;
each of them presents a peak, but at a different position,
suggesting the presence of two distinct cores
which will be referred to  as C17 and C19.
The separation between C17 and C19 is $\sim$7\asec\ in the
NS direction, with C17 (the most intense)
slightly to the north of the water maser position and C19 (the weaker one)
to the south of it.

%
%

\begin{figure}
\centerline{\psfig{figure=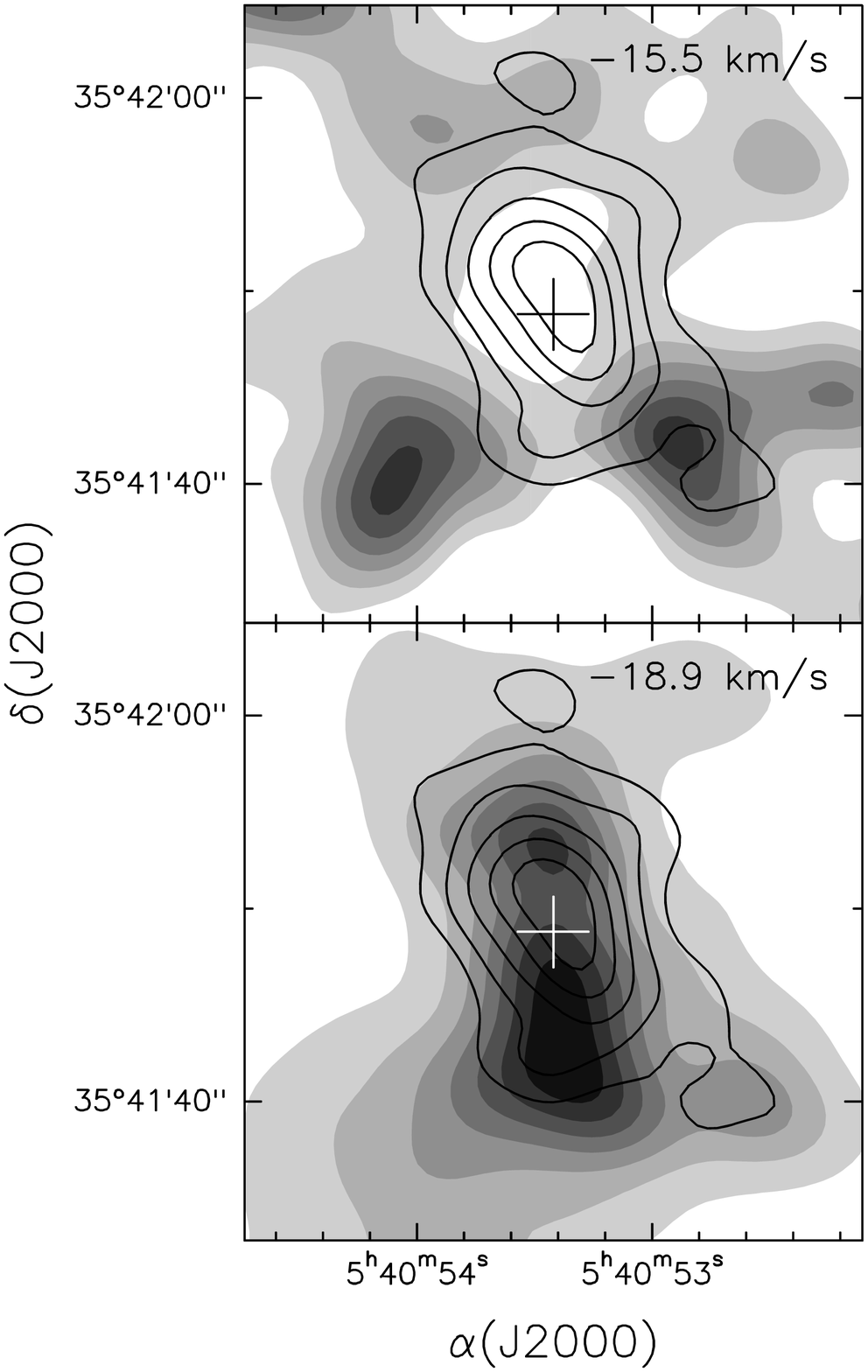,width=8.8cm}}
\caption[]{Gray scale maps of the HCO$^+$ emission at --15.5 \kms\ (top)
and --18.9 \kms\ (bottom). The contours reproduce the
CH$_3$CN of  Fig.~\ref{hcoint}.
The water maser position is indicated
by a cross. 
Gray levels are from 0.029 to 0.177 in steps of 0.029 Jy/beam
in the top panel and from 0.107 to 1.606 in steps of 0.249 Jy/beam
in the bottom panel.
}
\label{dip}
\end{figure}

Finally, to establish if the dip on the red side of the
profile is due to infalling material, we have produced two maps from
channels symmetrically
placed in velocity with respect to the peak  HCO$^+$ velocity (--17.2 \kms),
i.e. at  --15.5 \kms\ (the velocity of the dip) and --18.9 \kms\ (the 
symmetric velocity). The two maps are shown in  Fig.~\ref{dip} in grey scale,
overlaid on the CH$_3$CN map (contours), which traces the densest
part of the molecular core.
The dip clearly coincides with the peak of the CH$_3$CN emission,
while at the symmetric velocity there is no decrease of HCO$^+$ emission
at the position of the CH$_3$CN peak (apart from some complication due to
the presence of the  HCO$^+$ C19  peak discussed above). 
This indicates that the dip is a signature of infall in the core.

The velocity of the infalling gas can be estimated using the
model of Myers et al.~(\cite{myers}) and applying their Eq.~(9),
which for the case of $T_{\rm D}$ (the brightness temperature of the dip)
much smaller than $T_{\rm BD}$ (the height of the blue peak above the dip)
and $T_{\rm RD}$ (the  height of the red peak above the dip) simplifies to:
\begin{equation}
V_{\rm in} \simeq \frac{\sigma^2}{v_{\rm red}-v_{\rm blue}}
		  \ln\left(\frac{T_{\rm BD}}{T_{\rm RD}}\right)
\end{equation}
where $\sigma$ is the velocity dispersion of the line
 and ($v_{\rm red}$--$v_{\rm blue}$) is
the difference between the velocities of the red and blue peaks. 
From  Fig.~\ref{profiles}, $T_{\rm BD}$=2.4~K, $T_{\rm RD}$=0.12~K, 
$v_{\rm red}$-$v_{\rm blue}$=4~\kms\ and from Fig.~\ref{pvpdb}
$\sigma$=2.6 \kms.
The derived $V_{\rm in}$ is $\sim$ 5 \kms.

Using $V_{\rm in}$ and making plausible assumptions on 
the radius ($R_{\rm in}$) to which $V_{\rm in}$ refers and
the corresponding density ($\rho=3 M/4 \pi R_{\rm in}^3$),
one can estimate the mass infall rate:
\begin{equation}
\dot{M}_{\rm in} = 4 \pi R_{\rm in}^2 \rho V_{\rm in} =
		   3 \frac{M V_{\rm in}}{R_{\rm in}}
\end{equation}
For $M$ and $R_{\rm in}$ we have used the values derived from the mm continuum
observations, given in Table~\ref{massethin}. The derived value
is $\dot{M}_{\rm in} \simeq 10^{-2}~M_{\odot}$~y$^{-1}$.

%
%

\subsubsection{CH$_3$CN}
\label{ch3cn}

\begin{figure}
\centerline{\psfig{figure=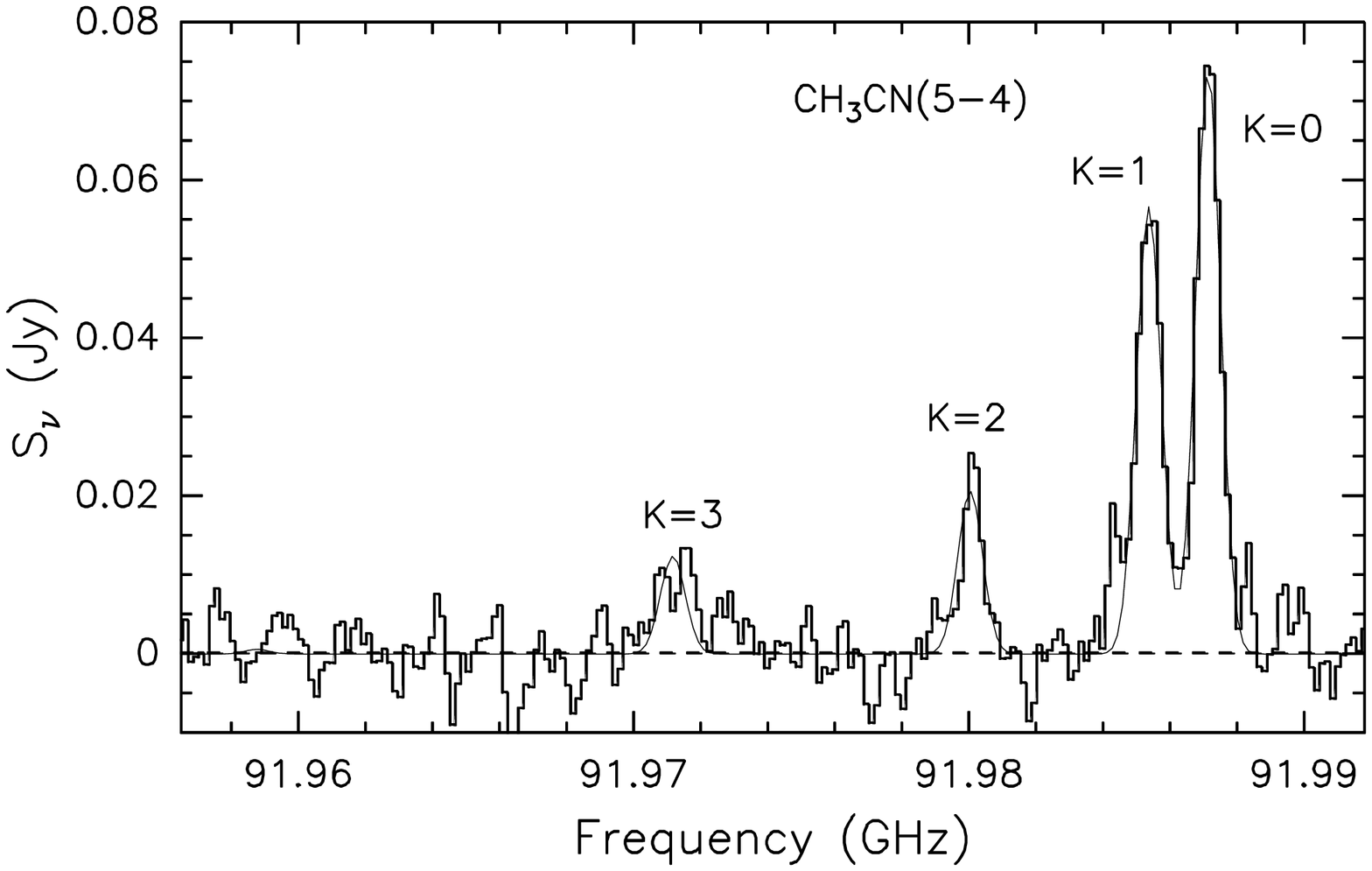,width=8.8cm,angle=-90}}
\caption[]{ The CH$_3$CN line profile of the emission at the peak position
averaged  over  the synthesized  beam.
The thin line is a gaussian  fit to the observed lines.
}
\label{ch3cnp}
\end{figure}

The CH$_3$CN(5--4) $K$=1 map averaged 
from --20.63 \kms\ to --15.03 \kms\ is shown in Fig.~\ref{hcoint}b.
The diameter of the emitting region  is $\sim$7\asec\ after beam 
deconvolution.
The two maps in the velocity
intervals from --17.58 to --15.54 \kms\ and from --19.10 to --17.58 \kms\ 
are shown in Fig.~\ref{peaks}c-d.
The CH$_3$CN line profile obtained by averaging over the synthesized  beam 
is shown in Fig.~\ref{ch3cnp}.

%
%

The two bipolar outflows seen in the HCO$^+$ line are not detected
in the CH$_3$CN line  since  it is not
possible to use the stronger $K$=0 and $K$=1 transitions because of 
blending  of the lines, while the higher order transitions,  
where blending is less severe, are too weak. Also, CH$_3$CN
is a tracer of high density gas and not of bipolar outflows.

As in the HCO$^+$ line the CH$_3$CN emission 
shows two distinct velocity peaks (once allowance is made
for the lower intensity of the line, see Fig.~\ref{peaks}c-d). Correspondingly, 
the two components C17 and C19 can be found in the same velocity intervals and
at the same position as those observed in the HCO$^+$ line. 

\subsubsection{C$^{34}$S, H$_2$CS and SO$_2$}
\label{mmlines}

\begin{figure}
\centerline{\psfig{figure=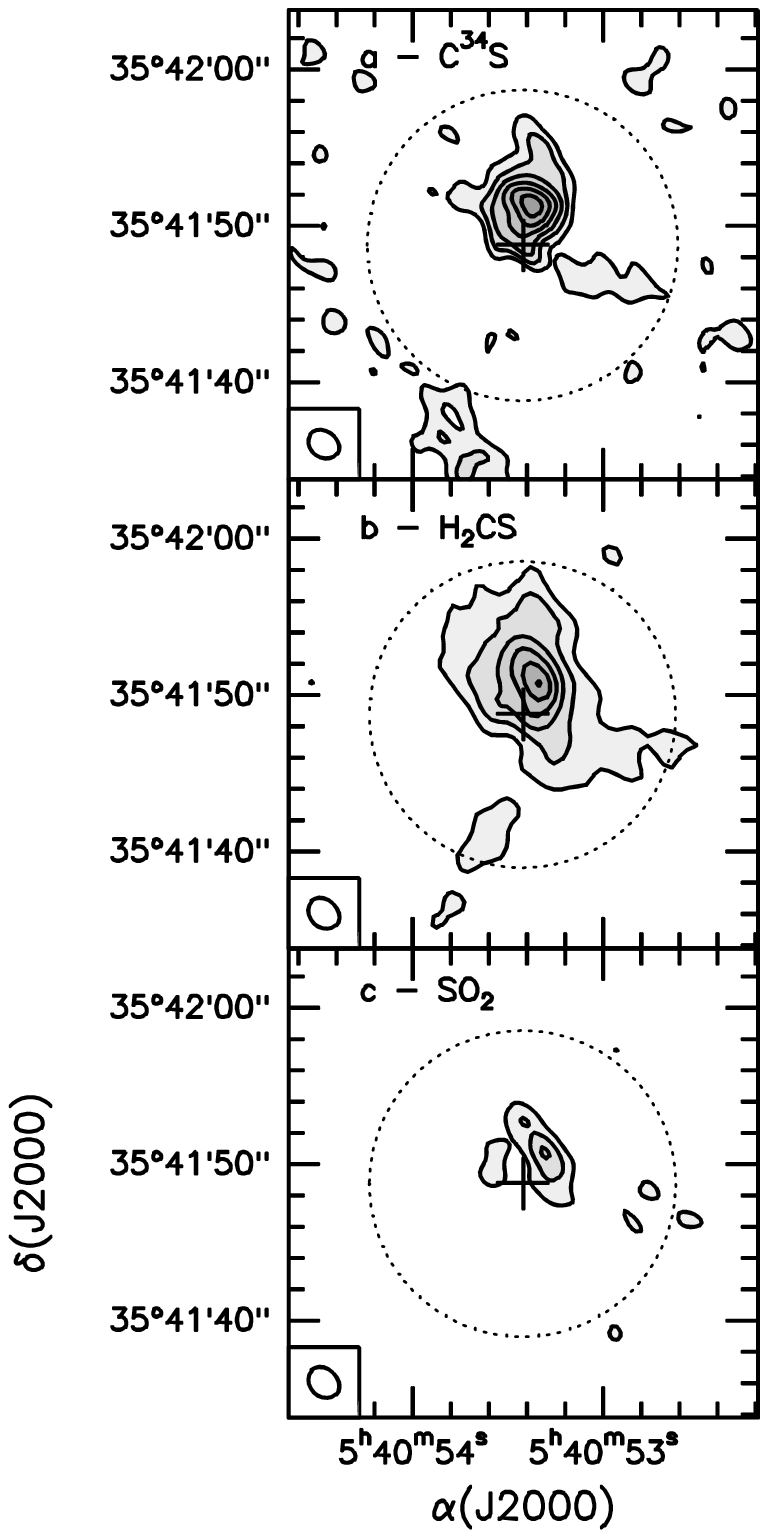,width=8.8cm}}
\caption[]{ The C$^{34}$S (a), H$_2$CS (b) and SO$_2$ (c) 
maps of the velocity averaged emission.
The water maser position is indicated
by a cross. The scale is identical to that of the 1.2  mm map
in Fig.~\ref{mmmap}. The dotted circle defines the primary beam HPW.
The box in the lower left shows  the synthesized HPBW.
Contour levels range from 6 to 36 by 6 mJy/beam for C$^{34}$S,
from 15 to 175 by 30 mJy/beam for H$_2$CS and from 25 to 75 by 25 mJy/beam.
for SO$_2$.
}
\label{1.2mmlines}
\end{figure}

The maps in the 1.2 mm lines of C$^{34}$S, H$_2$CS and SO$_2$
(see Table~\ref{tfreq} for the frequency set-ups)
are shown in Fig.~\ref{1.2mmlines}, not corrected for primary beam attenuation.
All of these confirm that the molecular emission peaks at the
position of the water maser. 
The diameter of the core  in the C$^{34}$S line is $\sim$3\farcs5.

Most of the H$_2$CS emission appears to be centred at $\sim-18$~\kms\
and clearly traces C17.
The H$_2$CS map shows 
also  a tail (at a 3$\sigma$ level) in the same direction and
position as  the jet observed in the continuum  at 3.3 and 1.2 mm, with  
emission concentrated around --19 \kms\ and partly overlapping with C19.

SO$_2$ is very faint, so no velocity structure can be derived from
our data. It is concentrated towards C17, but it can be noted that
its emission seems to be stretched in the NE--SW direction, roughly 
along the NE--SW outflow.

\begin{figure}
\centerline{\psfig{figure=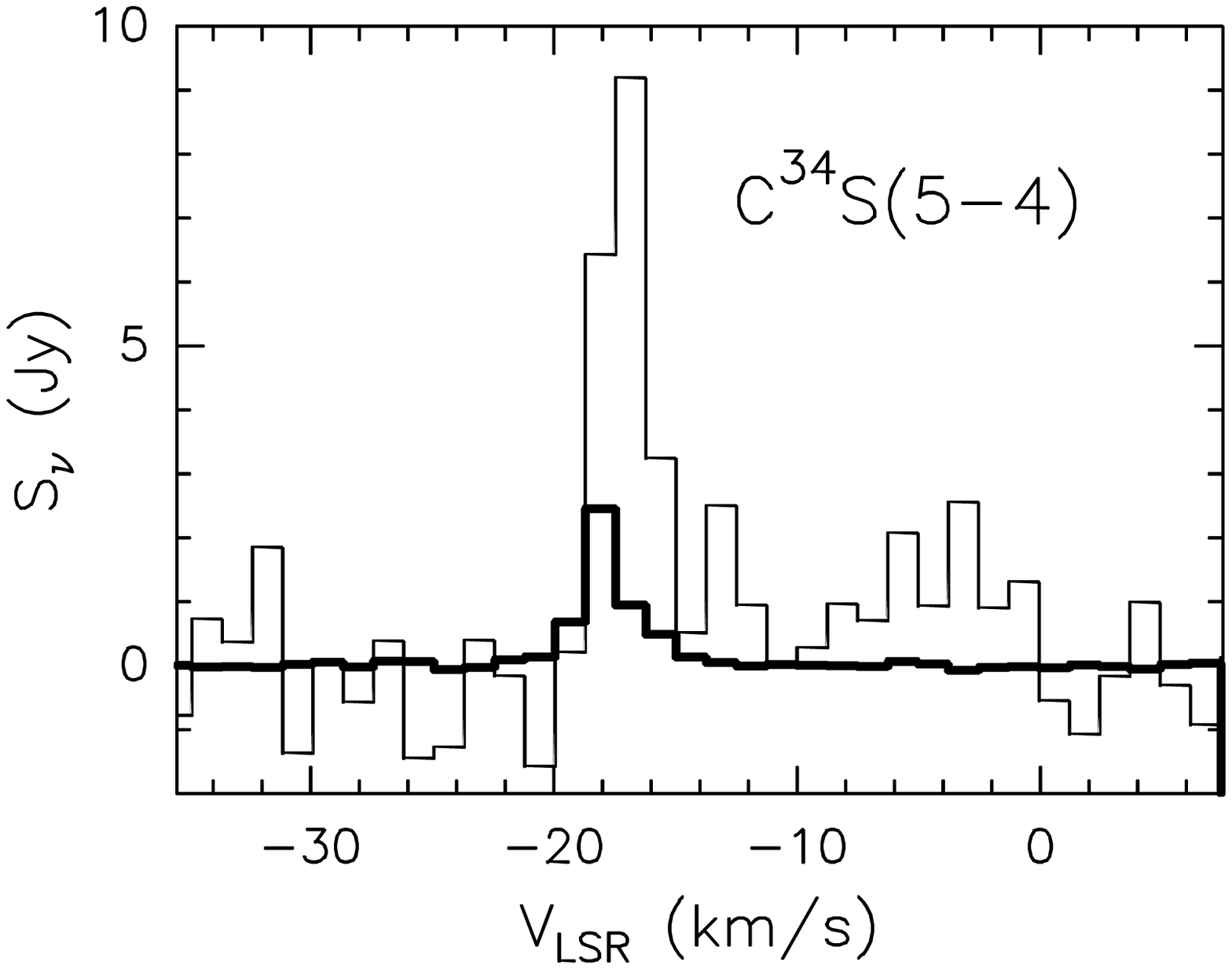,width=8.8cm}}
\caption[]{Comparison between the  C$^{34}$S profile observed at
Pico Veleta with a resolution of 10\asec\ (thin line) and that observed at
Plateau de Bure over a similar area with a resolution of $\sim$2\asec\
(thick line)
}
\label{pvpdb}
\end{figure}

The comparison between the C$^{34}$S line profile
observed at Pico Veleta and that at  Plateau de Bure is shown in 
Fig.~\ref{pvpdb}.
Most of the C$^{34}$S flux comes from structures resolved out in the
interferometric observations. Hence,  the molecular core detected in 
our high resolution 
observations represents just a small (high density)
part of the molecular cloud present in the S235A-B region.

%
%
%
%

%
%

\subsection{James Clerk Maxwell Telescope -- SCUBA}
\label{jcmt}

The submillimetre observations were carried out with the Submillimetre
Common-User Bolometer Array (SCUBA; Holland et al.~\cite{scuba}) on the James
Clerk Maxwell Telescope on Mauna Kea, Hawaii on October 24, 2000 and
were obtained from the public archive at the CADC. SCUBA consists of a
short wavelength array containing 91 pixels and a long wavelength
array containing 37 pixels arranged in a close-packed hexagon. Both
arrays have approximately the same field-of-view (2.3 arcmin) and can
be used simultaneously. These observations used the 450- and
850-micron filters and were taken using the ``Emerson II'' scanning
technique (Emerson~\cite{emerson}; Jenness et al.~\cite{jhcld}) 
where the telescope is
scanned in Nasmyth coordinates (in order to generate fully-sampled
images) at 24\asec/sec whilst the secondary is chopped in fixed
directions on the sky. Six chop configurations were used: 30\arcsec,
44\arcsec, and 68\asec\ 
in right ascension and similarly in declination. A single map
was taken for each chop configuration, covering a region of 3 by 10
arcmin and took approximately 35 minutes. CRL~618 was used for flux
calibration (Jenness et al.~\cite{jsaejr}) with minor corrections applied
using pointing observations on 0552+398 straddling the science
observations. The zenith atmospheric opacity was obtained using a fit
to the CSO 225~GHz tipping radiometer data as described in Archibald
et al.~(\cite{ajh}) and was approximately 0.07 (225~GHz). The data were
reduced using the standard SCUBA User Reduction Facility (SURF;
Jenness \& Lightfoot~\cite{JL98}) software.
The half power beam widths (HPBWs)
are 14\asec\ at 850 and 7.5\asec\ at 450 ${\mu}$m.

%
%
\begin{figure*}
\centerline{\psfig{figure=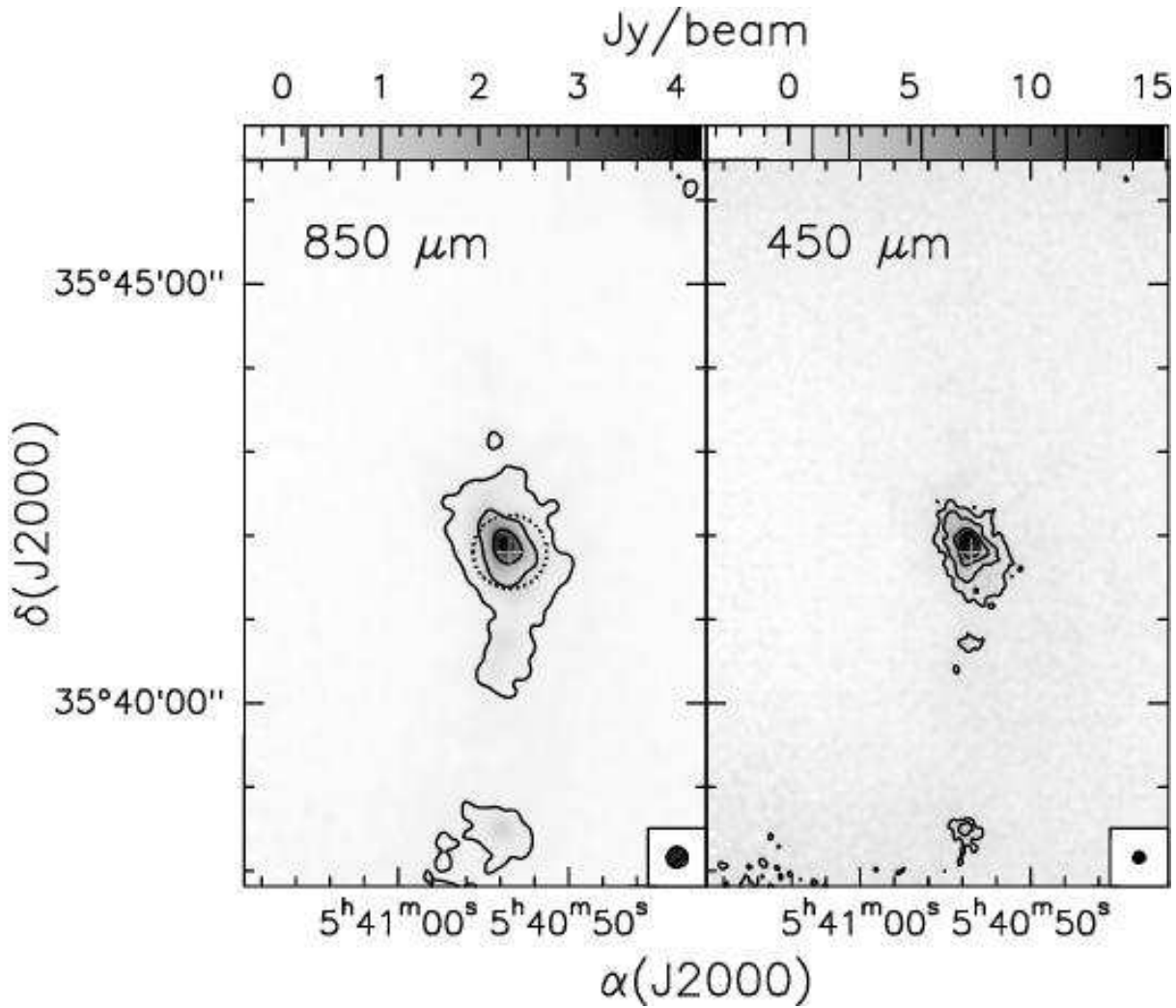,width=16cm}}
\caption[]{ The SCUBA maps at 850 ${\mu}$m and 450 ${\mu}$m.
The water maser position is indicated
by a cross. The box in the lower right  shows the HPBW.
The dotted circle in the 850 ${\mu}$m map defines the HPW of the 
primary beam at 3.3 mm. The values of the contours  are indicated 
in the grey wedges on top of each map. The source  at the bottom of the
two maps is a real detection and is related to the \HII\ region S235C
(Israel \& Felli~\cite{IF}).
 
}
\label{scuba1}
\end{figure*}

The SCUBA maps at 850 ${\mu}$m and 450 ${\mu}$m are shown in
Fig.~\ref{scuba1}.
At both wavelengths  the peak coincides with the water maser.

\begin{figure}
\centerline{\psfig{figure=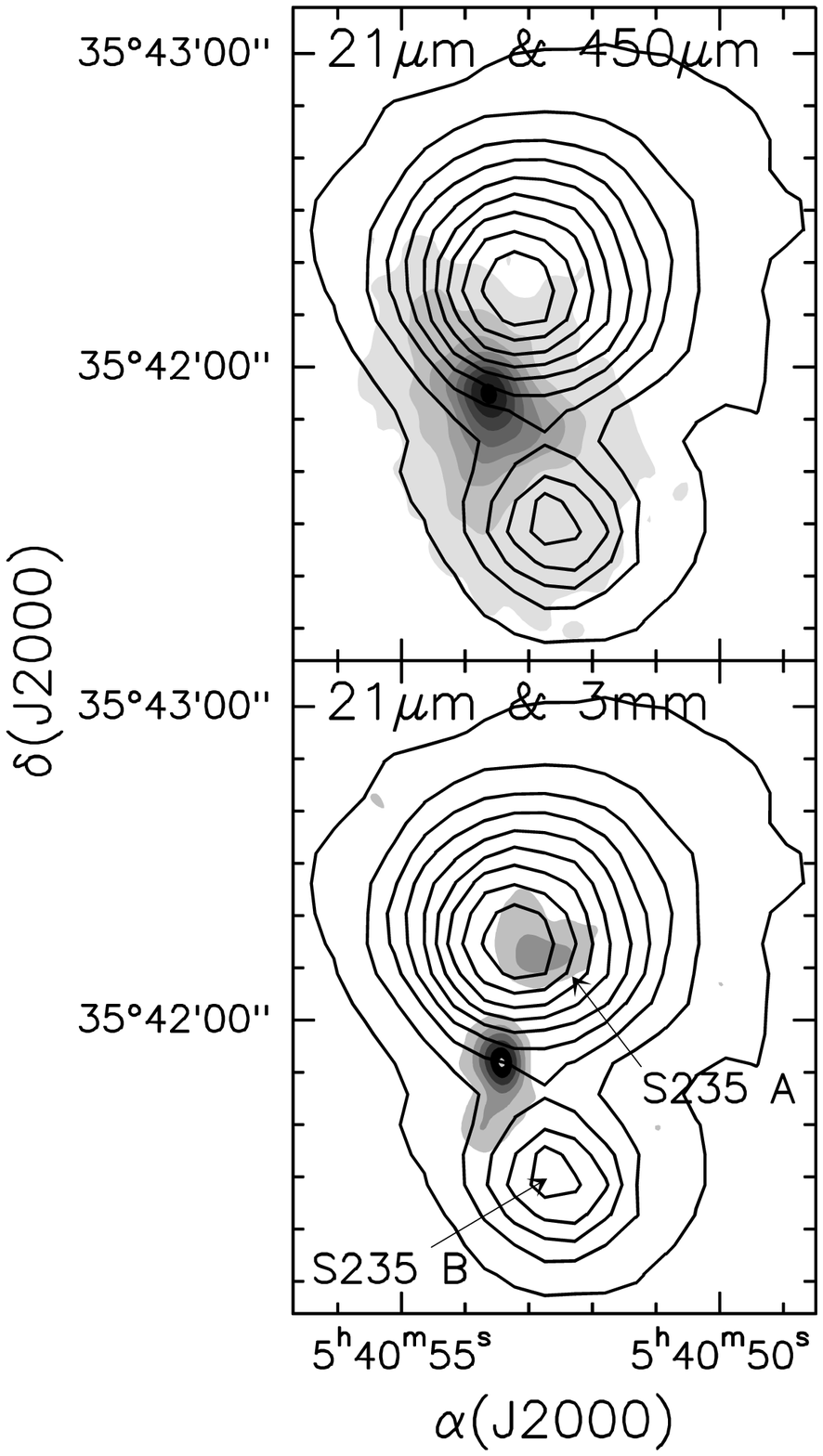,width=8.8cm}}
\caption[]{Top panel: overlay of the 21 ${\mu}$m maps of MSX (contours)
with the 450 ${\mu}$m emission (grey scale). Bottom panel: 
overlay of the 21 ${\mu}$m maps of MSX (contours) with the 3.3 mm emission
(grey scale). Contours range from 10$^{-5}$ to  1.9 10$^{-4}$ in steps of  
2 10$^{-5}$ Watt m$^{-2}$ sterad$^{-1}$.
}
\label{mmscuba}
\end{figure}

%
To show the relationship between the mm and sub-mm  peaks
with the mid-IR emission we have overlaid in Fig.~\ref{mmscuba}
the 450 ${\mu}$m and 3.3 mm maps with the 21 ${\mu}$m map
of  the Midcourse Space Experiment (MSX, resolution 20\asec). 
It is clear that the 21 ${\mu}$m
emission comes predominantly from the S235A and S235B nebulosities, 
with  very little contribution (if any) from the YSO associated 
with the water maser, the mm core  and the 450 ${\mu}$m core.

%
%

\begin{figure}
\centerline{\psfig{figure=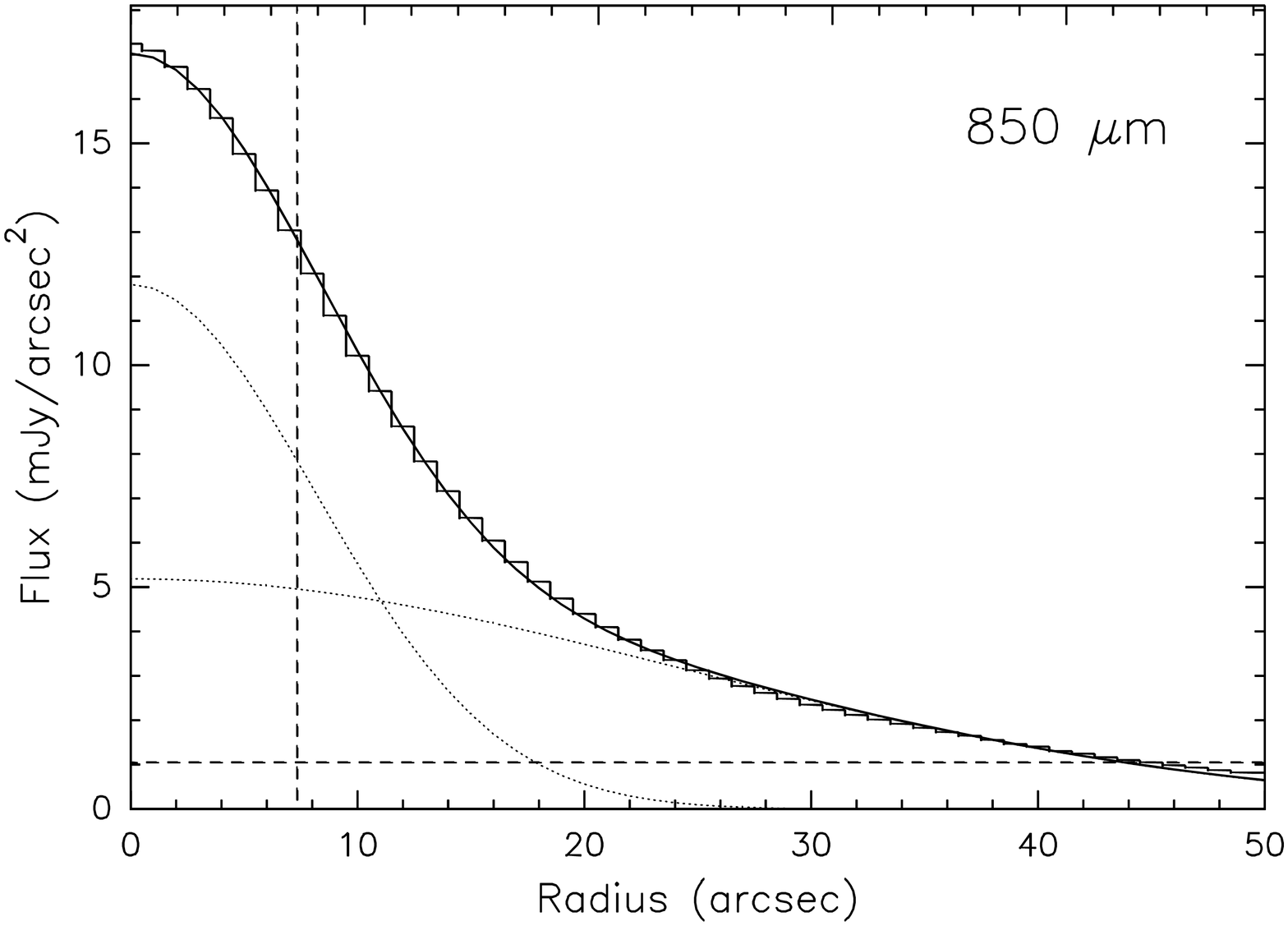,width=8.8cm,angle=-90}}
\caption[]{ The profile of the emission at  850 ${\mu}$m
integrated over circular rings of increasing radii. 
The two gaussians (convolved with the beam) 
whose sum  best fits the observed profile are
represented by the dotted lines. The vertical dashed line
represents the HPBW and the horizontal dashed line the 3 $\sigma$
noise level.
}
\label{strip}
\end{figure}

The two SCUBA maps show a compact source at the centre, surrounded by a more
extended halo. This is more evident at 850 ${\mu}$m.
To separate the  two components,
a radial profile of the 
emission was obtained at the two wavelengths 
by measuring the flux density contained 
in circular rings with increasing radii departing from the peak 
position.  The profile at 850~${\mu}$m is shown in Fig.~\ref{strip},
together with  its decomposition into two gaussians of half power width (HPW)
equal to  12\asec\ and 56\asec. 
The corresponding flux densities for the compact source 
and  the entire source 
are 5  Jy and 24 Jy at 850 ${\mu}$m and 11 Jy and 84 Jy at 450 ${\mu}$m.

%
%

\begin{figure}
\centerline{\psfig{figure=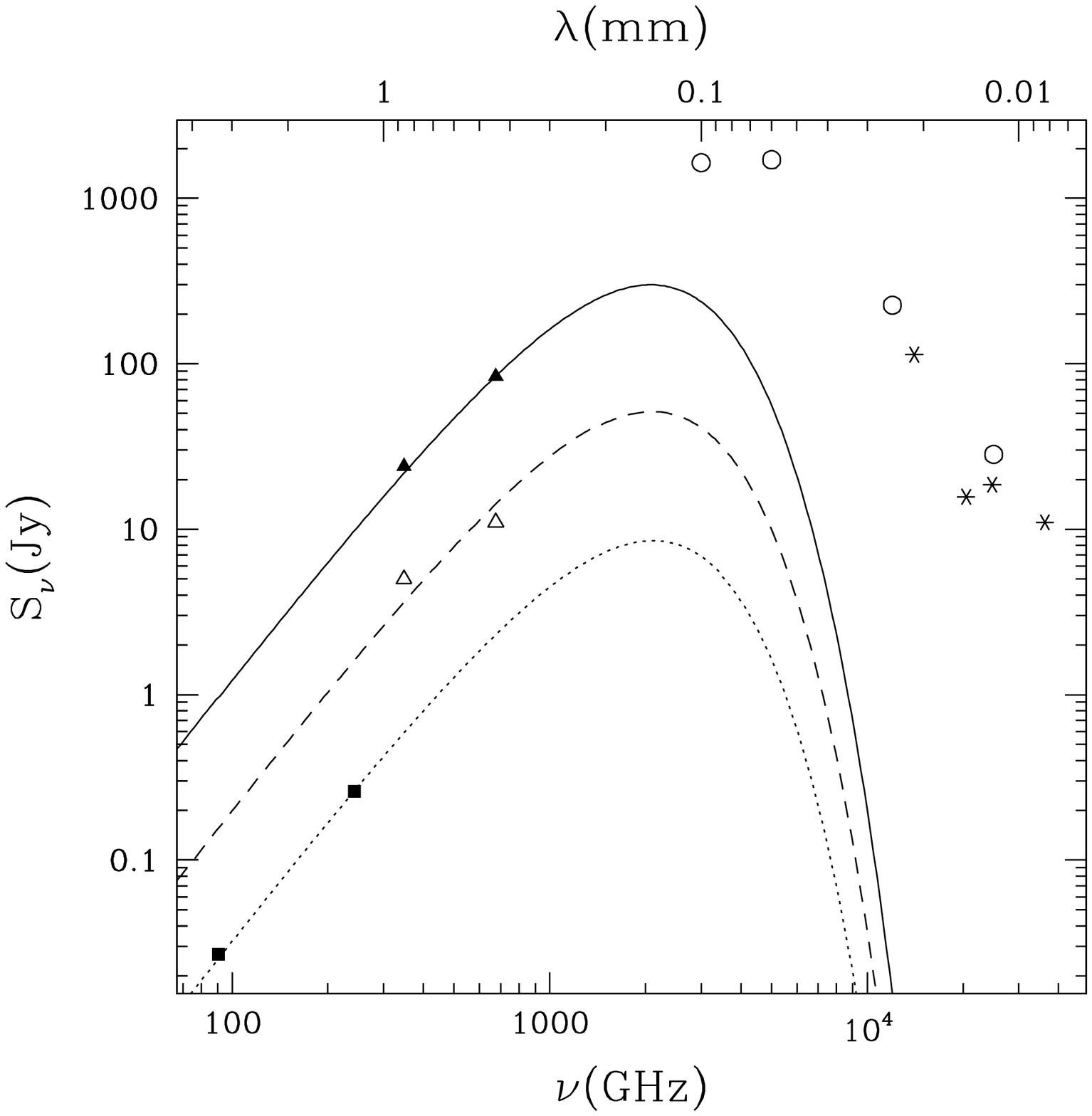,width=8.8cm}}
\caption[]{ The SED of the region around the water maser.
The two black squares are the total 3.3 and 1.2 mm flux densities,
the filled triangles are  the total flux density and  the empty triangles
the 
flux density of the compact core  
at 850 and 450 ${\mu}$m, the four circles are the IRAS flux densities,
the four asterisks  the MSX  flux densities.
The three lines denote grey-body fits to the mm and sub-mm data
discussed in the text.
}
\label{sed}
\end{figure}

The Spectral Energy Distribution (SED) from 3.3 mm to 8 ${\mu}$m
is shown in Fig.~\ref{sed}. In view of the different resolutions
of the MSX and IRAS data with respect to the present observations 
and the results of the overlays of  Fig.~\ref{mmscuba}, 
the IRAS and MSX data
refer essentially to the S235A-B nebulosities
and will not be considered further.
The fits to our data will be explained in Sect.~\ref{score}.

%
%

\begin{figure}
\centerline{\psfig{figure=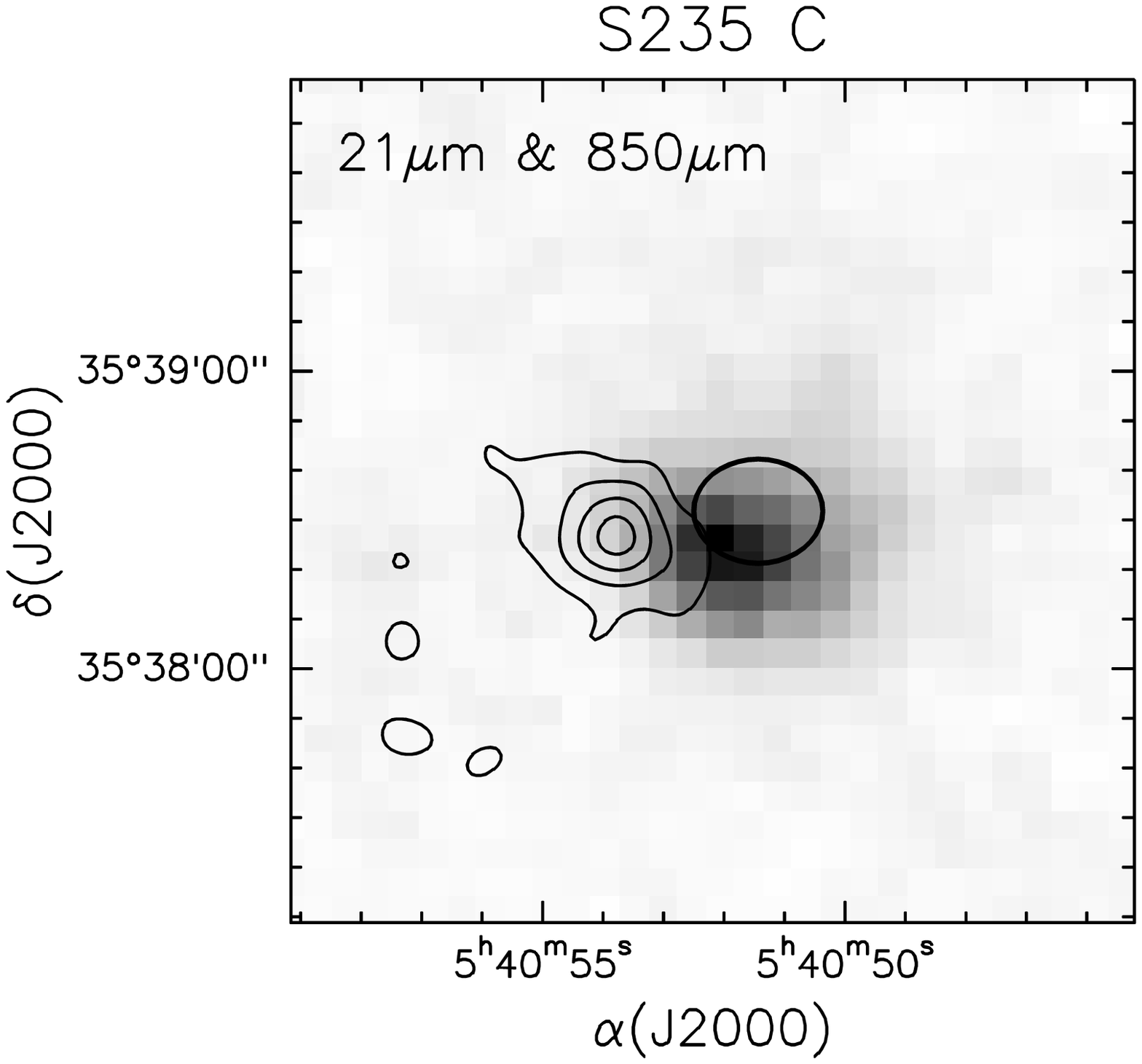,width=8.8cm}}
\caption[]{Overlay of the 850 ${\mu}$m map of the region around  S235C
(contours) with the 21 ${\mu}$m  map from MSX (grey scale). The ellipse
schematically indicates the position and size of the 6 cm radio continuum 
emission from the S235C \HII\ region (from Israel \& Felli~\cite{IF}).
}
\label{S235C}
\end{figure}

At the southern edge of the maps in Fig.~\ref{scuba1}, well outside the primary
beam of
the interferometric observations,  another sub-mm source is present.
This source coincides with S235C. 
The radio  continuum  flux densities of S235C at 21 and 6 cm 
are 70 and 60 mJy (Israel \& Felli~\cite{IF}),
and indicate that S235C is an \HII\ region,  presumably ionized by a B0.5 star.
In the same area there are also IRAS\,05375+3536 and an MSX source.
The flux densities at 850 ${\mu}$m  and 450 ${\mu}$m are 1.0 $\pm$ 0.1 Jy and
3.5 $\pm$ 0.5 Jy, respectively. This new result clearly indicates the presence
of dust associated with  S235C.
In Fig.~\ref{S235C} we show an overlay of the  850 ${\mu}$m map with the
21 ${\mu}$m image derived from MSX.  The position and size of the 6 cm 
radio continuum source are marked with an ellipse.
The not perfect match between the 850 ${\mu}$m emission  and the 
21  ${\mu}$m and radio continuum emission shows that two close-by YSOs 
in different evolutionary stages may be present in this area: the S235C 
\HII\ region and the 850 ${\mu}$m source, 
representing most probably  the emission from
dust associated with a YSO in an earlier evolutionary phase.
The  situation  is 
very similar to that occurring in the S235A-B complex.
However, given the lack of interferometric information on S235C, 
no further discussion of this star forming region is possible.

%
%

\section{Discussion}
\label{discussion}

\subsection{The molecular core}
\label{score}

One of the main results of the present work is the confirmation
that a high density molecular and dusty  core, part
of the larger scale molecular complex detected in the CO lines
(see e.g the $^{13}$CO observations of Cesaroni et al.~\cite{CFW}), 
is located close to the water maser. This morphology is typical
of the earliest stages in the evolution of a YSO and proves that
a new episode of star formation is occurring near the water maser, while
other star formation events like those associated with S235A and S235B 
occurred in the past and in S235A  have had sufficient time to create a well
developed \HII\ region.
  
\begin{figure}
\centerline{\psfig{figure=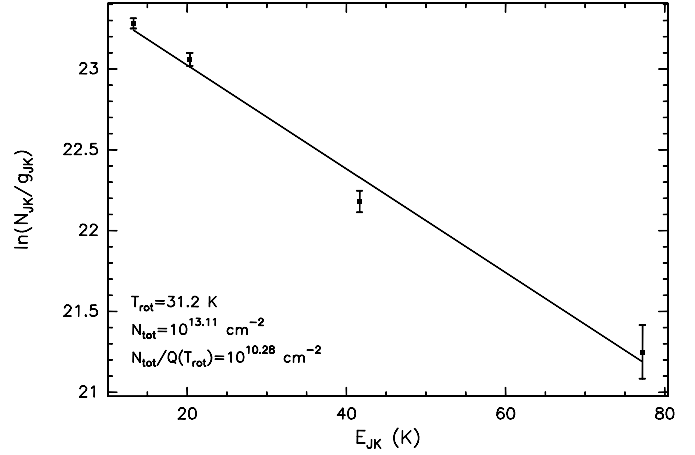,width=8.8cm,angle=-90}}
\caption[]{Boltzman plot for the CH$_3$CN(5--4), $K$=0,1,2,3 lines.
The column densities are source averaged. The straight line represents
a least squares fit to the data.
}
\label{bolt}
\end{figure}

The temperature of the molecular core can be  derived from
the CH$_3$CN(5--4), $K$=0,1,2,3 lines measured towards the  peak position by
means of the rotation diagram method (Boltzman plot), which assumes that
all energy levels of CH$_3$CN are populated according to
local thermodynamic equilibrium (LTE) with a single excitation 
temperature. The results of the fit are shown in Fig.~\ref{bolt}
and give a temperature of $\sim$30 K. This value is smaller than 
the 43~K peak main beam brightness temperature of the $^{13}$CO(2 -- 1)
line (Felli et al.~\cite{FTVW}) and suggests that the optically thick CO 
emission of the larger scale molecular
cloud  might be heated from outside by the radiation 
of the star  exciting the S235A \HII\ region, while  the molecular
core around the water maser is shielded from the \HII\ region radiation 
and excited only by the YSOs placed inside.

As noted in Sect.~\ref{mmlines}, 
the molecular core is also traced by H$_{2}$CS and SO$_{2}$,
which are predominantly 
found towards hot cores (Hatchell et al.~\cite{htmm}) or the
envelopes of massive stars (van der Tak et al.~\cite{tbbd}). 
In particular, van
der Tak et al.~(\cite{tbbd}) found that H$_{2}$CS arises from regions with
excitation temperatures of $\sim 50$ K, which is not too different from 
our  estimated temperature, whereas SO$_{2}$ shows a higher excitation
temperature ($\sim 100$ K). 

A large fraction of SO$_{2}$ is usually found
in high-velocity gas (van
der Tak et al.~\cite{tbbd}); this aspect  might be consistent with the
NE--SW elongated shape of the SO$_{2}$ emission along the direction of the 
NE--SW outflow.
Indeed, sulphur-bearing molecule
production is believed to originate either from grain evaporation in the
dense hot environment around protostars, or from shocked gas (Hatchell et
al.~\cite{htmm}; Bachiller et al.~\cite{bgkt}; van der Tak et al.~\cite{tbbd}).
From our observations with  only one transition per molecule, 
we cannot state whether they mostly arise from warm gas or from
shocked gas around the protostar, 
although part of
the emission of H$_{2}$CS is clearly related with the outflows (hence,
with shocked gas). 

%
%

To derive the  luminosity of the YSOs and the mass of the molecular cloud
we use the mm and sub-mm flux densities.
Figure~\ref{sed} shows  grey-body fits to our data, with the
assumption of constant density and constant temperature in the
molecular cloud. The best fit is  
obtained using a dust opacity $\kappa_{\nu} \propto \nu^{0.5}$
and the  temperature  derived from the CH$_3$CN lines.
Larger dust opacities ($\kappa_{\nu} \propto \nu^{1 - 2}$)
produce worse fits to the mm and sub-mm data.

The lower dotted curve  in Fig.~\ref{sed} shows a fit 
to the mm data alone. In this fit no distinction was made
between core  and jet and total flux densities were
used. It should also be noted that the 1.2 mm observations
may resolve out, and consequently miss, 
%
some of the extended emission. With these premises, the  
luminosity is  24 \lsol\
and the total mass  16 \msol.
These values of mass and luminosity
might refer to a possible disk around the YSO
and  should be considered as an upper limit to the true disk
values since contamination from the dust in the core
that surrounds the disk might occur.
The middle dashed curve is a fit to the two sub-mm flux densities
of the  compact source (see Sect.~\ref{jcmt}). The luminosity is 160 \lsol\ and
the mass 101 \msol. 
These values should refer to the densest part of the molecular
core.  Finally, the upper full line is a fit 
to the two total sub-mm flux densities (compact source + halo) and gives
a luminosity of 920 \lsol\ and a mass of 690 \msol.
These values are representative of the total luminosity of the YSO (or cluster 
of YSOs) and the total mass of the molecular core in which the YSO(s) is (are) 
harboured.

There are two other ways to derive the mass of the molecular core: the
first (virial mass) is from the width of the line (from gaussian fits) 
and the source size. We used the relation of MacLaren et al.~(\cite{mrw})
for a homogeneous sphere.  The second uses the
intensity of optically thin lines and assumes LTE and a given
molecular abundance with respect to H$_2$. 
The derived masses  are given in Table~\ref{massevir} and Table~\ref{massethin},
respectively.  The values obtained from HCO$^+$
which refer to a resolved source of 0.1 pc in size,  
are similar and consistent, within the
approximation of the present estimates, with the mass of the dense
core derived from the fit to the JCMT flux densities. Those
derived from C$^{34}$S (similar for the two methods) refer to a smaller
and probably denser central component and are consistent with the 
disk mass obtained from the fit to the mm flux densities.
Finally, the CH$_3$CN estimates are not fully consistent with each other, very
likely due to the large uncertainty in the molecular abundance assumed.

%
%
\begin{table}
\caption[]{Diameter, velocity width and virial masses} 
\label{massevir}
\begin{tabular}{cccc}\hline
Line  & Diameter  &  $\Delta V$ & Virial Mass \\
 & (pc) &  (km s$^{-1}$) & ($M_{\sun}$) \\ 
\hline
HCO$^{+}$(C17) $^{a}$ & 0.1 &  2.4 & 58 \\
HCO$^{+}$(C19) $^{a}$ & 0.08 &  3.1 & 77 \\
C$^{34}$S$^{a}$ & 0.03 &  1.49 & 7 \\
CH$_{3}$CN & 0.06  &  2.85 & 53 \\
\hline
\end{tabular}

\vspace*{1mm}
{$^{a}$ Line wings were excluded from fits.}
\end{table}

%
%
\begin{table}
\caption[]{Diameter and mass of the molecular core}
\label{massethin}
\begin{tabular}{ccc}\hline
Tracer  & Diameter  & Mass \\
 & (pc) & ($M_{\sun}$) \\
\hline
HCO$^{+}$ (C17)$^{a}$ & 0.1 &  31 \\
HCO$^{+}$ (C19)$^{a}$ & 0.08  & 10 \\
C$^{34}$S$^{a}$ & 0.03 & 7 \\
CH$_3$CN$^a$ & 0.06 & 2.3 \\
mm continuum & 0.054 & 16 \\
\hline
\end{tabular}

\vspace*{1mm}
$^{a}$ T$_{\rm ex}$=30 K, 
[HCO$^{+}$]/[H$_{2}$]$=10^{-9}$,
[C$^{34}$S]/[H$_{2}$]$=10^{-10}$,
[CH$_3$CN]/[H$_{2}$]$=5\times10^{-10}$.
\end{table}

\subsection{The YSOs}

In this section we examine the possible implications that
can be derived for the YSOs predicted to exist in the molecular core
from  the NIR observations of Felli et al.~(\cite{FTVW}).
  
The NIR source closest to the water maser 
is M1 (Felli et al.~\cite{FTVW}) and it is also the reddest one in the 
entire field.
M1 is the stellar object with the largest near-IR excess
in the region and it was
proposed by Felli et al.~(\cite{FTVW}) as the powering source of 
the water maser.
As shown in Fig.~\ref{hcnk} it is located
$\sim 5\arcsec$ SSE of the maser position and only $\sim 2 \arcsec$
east of C19 and it lies  
projected on the thermal jet. However, its distance from  the 
C17 core ($\sim 10 \arcsec$) is  greater
than the possible uncertainty in the near-IR astrometry and makes
a direct correspondence of M1 with the YSO in the centre of C17 very
unlikely. M1 could still be associated with the YSO in C17 if it is
due to reflected radiation from the embedded star, as has been 
suggested to be the case for the (proto)star  IRAS\,20126+4104 
(Cesaroni et al.~\cite{cesa1}) and W75N (Moore et al~\cite{mmym};
Hunter et al.~\cite{htft}; Shepherd et al.~\cite{sts}), where the NIR
emission is not coincident with the location of the YSO.
This would imply the existence of a low density channel 
with lower extinction through which NIR photons emitted by the YSO 
can escape from the dense core.

We can check the nondetection of a  K band source in the C17 core
by estimating  the required stellar extinction A$_V$ and consequent 
column density, 
and comparing it with the column density  derived from  our mm observations.
An appropriate  lower limit to the K magnitude in the region 
around M1 is 15.  To derive  A$_{\rm V}$ 
we shall  assume the pre-main sequence (PMS) star
to have at least a mass of 6 $M_{\sun}$ (see Sects.~\ref{sdisk} 
and~\ref{natu}). Using the evolutionary tracks of
Palla \& Stahler~(\cite{ps}), a $5 \times 10^{5}$~yr old 
PMS star of such mass at
1.8 kpc would have K=10.9~mag, yielding $A_{\rm V}$=37~mag. The corresponding
column density is $N({\rm H}_2) =  3.6 \times 10^{22}$ cm$^{-2}$.
The {\it average} column density
for C17 is $3 \times 10^{22}$ cm$^{-2}$ from HCO$^{+}$ and
$3.4 \times 10^{22}$ cm$^{-2}$ from the mm dust emission. 
%
These values are mutually consistent and confirm that
a 6~$M_\odot$ YSO hidden inside the core
could not be observed in the K band.

Given the proximity of M1 to C19, there is still another
possibility to consider, namely
that M1 is instead coincident with a different   YSO
located in C19 and powering the thermal jet.
The column density  in C19 is lower than in C17
($1.4 \times 10^{22}$~cm$^{-2}$ from 
HCO$^{+}$), 
but could even be overestimated since no thermal mm 
continuum is found at the same position.
This would allow one to detect a lower mass star,
considering that the putative YSO in C19 must have a lower luminosity,
but no quantitative statement can be made at this stage.
In conclusion, it cannot be excluded that M1 may be the powering
source of the thermal jet and the NIR counterpart of  a YSO in C19.
Higher resolution,  deeper
NIR observations with more accurate astrometry
are needed to settle this issue.

\subsection{The molecular outflows}

The map of the high velocity HCO$^+$ gas (Fig.~\ref{outflow})  
shows that this is distributed in two outflows, both of them
originating approximately from the position of the water maser.
The longest, better defined  and more collimated one is in the NE--SW 
direction, with
little overlap between the red and blue lobes. The collimation is
rather high; using the water maser as centre position and the 
HPW of the lobes transverse to the maximum length we obtain an 
opening angle of the lobes of $\sim$15\degr.

\begin{figure}
\centerline{\psfig{figure=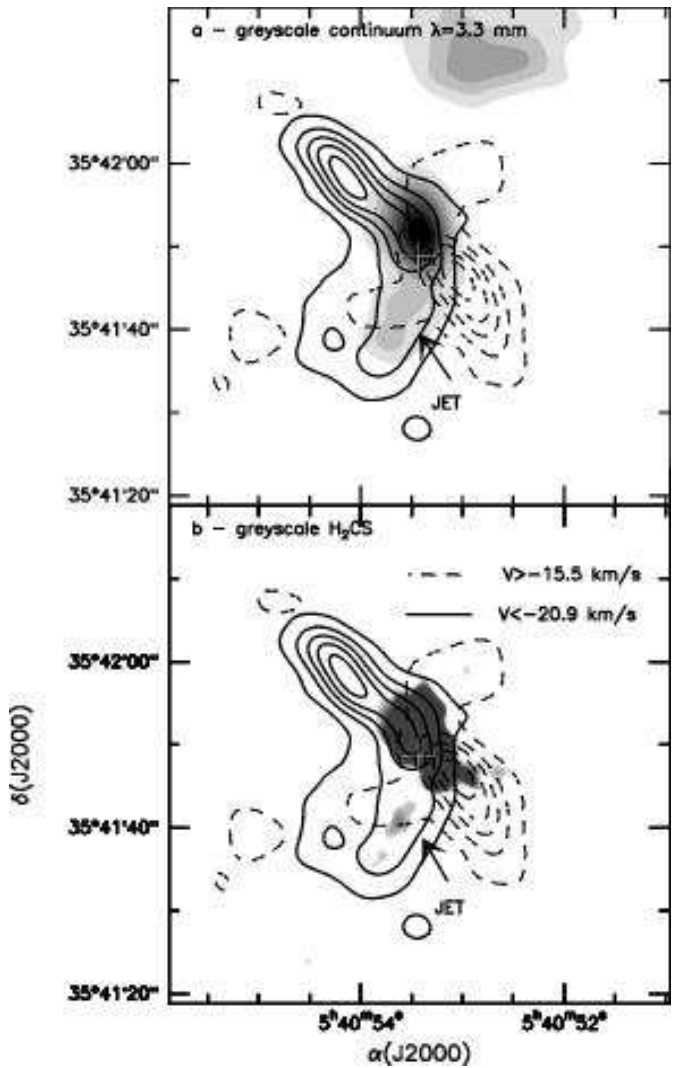,width=8.8cm}}
\caption[]{Top panel: overlay of the HCO$^+$ outflow (contours)
with the  3.3 mm continuum (grey scale). Bottom panel: overlay of 
the HCO$^+$ outflow (contours) with the H$_2$CS map 
(grey scale). The cross
marks the position of the water maser.
}
\label{outflow3h2cs}
\end{figure}

The other outflow is in the NNW--SSE direction, it is slightly 
less extended, less intense
and less collimated. Its southern blue lobe coincides with the thermal 
jet seen at 3.3 mm and with the H$_2$CS tail, as 
shown in Fig.~\ref{outflow3h2cs}.  
Sulphur-bearing molecules are expected (and found) in shocked gas within
outflows (Bachiller et al.~\cite{bgkt}), 
so the presence of H$_{2}$CS along the
NNW--SSE outflow  confirms that the molecular tail is part of the outflow.

%
%
\begin{figure}
\centerline{\psfig{figure=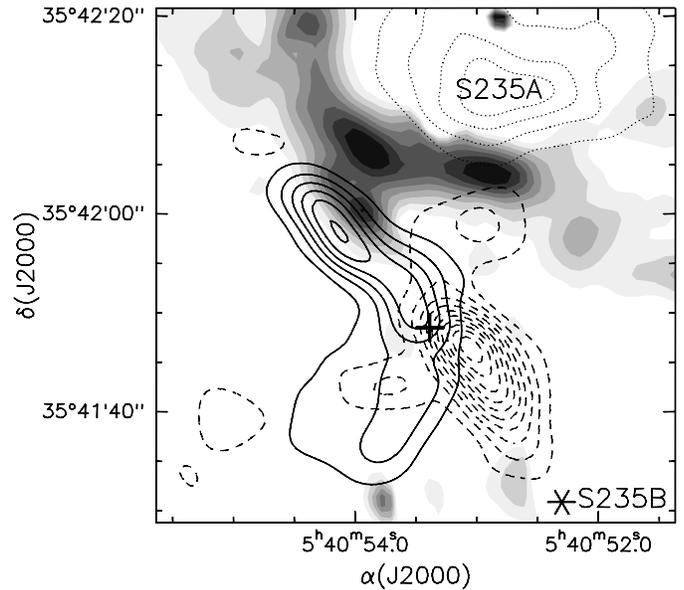,width=8.8cm}}
\caption[]{Overlay of the  H$_2$ emission (grey scale from
Felli et al.~\cite{FTVW}) with the  
HCO$^+$ outflow (contours).
The dotted-line contours delineate the 3.3 mm continuum emission from S235A.
The star marks the position of S235B, the cross that of the water maser.
}
\label{h2}
\end{figure}

Collimated outflows are often associated with shock excited H$_2$ emission
(see e.g. Cesaroni et al.~\cite{cesa2}). 
A comparison of Fig.~\ref{outflow} with  the H$_2$ emission
map from Felli et al.~(\cite{FTVW}) is shown  in Fig.~\ref{h2}.
Within the uncertainties of the respective astrometry, 
the northern blue lobe of the NE--SW outflow seems to coincide
with a distinct feature of the H$_2$ emission that clearly 
deviates from the H$_2$ half ring that surrounds the S235A \HII\ region.
There are three other possible associations in Fig.~\ref{h2} between 
H$_2$ features and the bipolar outflows, even though they 
should be checked with more accurate astrometry and more sensitive
H$_2$ observations: in the northern part, 
the red  lobe of the NNW--SSE outflow seems to end up in
a bright feature of the H$_2$ emission and, in the south, 
two weaker H$_2$ spots are located at the end of the blue lobe of 
the NNW--SSE outflow and of the red lobe of the NE--SW outflow. 

The possible implication is that  part of the  H$_2$ (the ring
around S235A) could be excited  in the photo dissociation region
at the interface
between the \HII\ region and the molecular cloud, while the H$_2$
associated with the outflows could be shock excited.  
Near-IR spectroscopy
of the H$_2$ features is needed to settle this point.

The parameters of the two outflows  were derived assuming [HCO$^+$]/[H$_2$]
= 10$^{-9}$, an excitation temperature of 30 K and optically thin 
emission, and are given in Table~\ref{toutflows}. 
The main uncertainty in these values comes from the artificial 
separation of the two outflows in the regions where they overlap.
All the derived parameters 
are smaller than those of the outflow found in IRAS\,20126+4140 
(Cesaroni et al.~\cite{cesa1}), as expected considering that 
IRAS\,20126+4140 has a bolometric luminosity about an order of magnitude 
greater. 

The maximum expansion velocities of the two lobes of the  NE--SW outflow
are not equal. The  blue lobe velocity reaches  
up to $\sim$15 \kms\ from the molecular core peak velocity, while that
of the red lobe is
at least twice as high. This explains the larger dynamical 
time scale found for the blue lobe, given the fact that the two lobes 
have almost equal sizes.

\begin{table*}
\caption[]{Physical parameters of HCO$^{+}$ outflows.} 
\label{toutflows}
\begin{tabular}{lccccc}\hline
 & Time scale & Mass &Mass loss rate & Momentum & Mech.\ Lum.\  \\
 & (yr) & ($M_{\sun}$) & ($M_{\sun}$ yr$^{-1}$) & ($M_{\sun}$ km s$^{-1}$) & ($L_{\sun}$) \\
\hline
NE--SW blue lobe &  14700 & 3 &  $2 \times 10^{-4}$ &  $ 26$ & $ 1.9$ \\
NE--SW red lobe & $\la 5400$ & 6 & $\ga  1.1 \times 10^{-3}$ & $\ga 65$ & $\ga 17$ \\
NNW--SSE blue lobe &  15000 & 3 & $ 2 \times 10^{-4}$ &  $ 13$ & $ 0.3$
 \\
NNW--SSE red lobe &  20000 & 1 & $ 5 \times 10^{-5}$ & $ 3$ & $ 0.05$
\\
\hline
\end{tabular}
\end{table*}

%
%

\subsection{The two velocity peaks}

\begin{figure}
\centerline{\psfig{figure=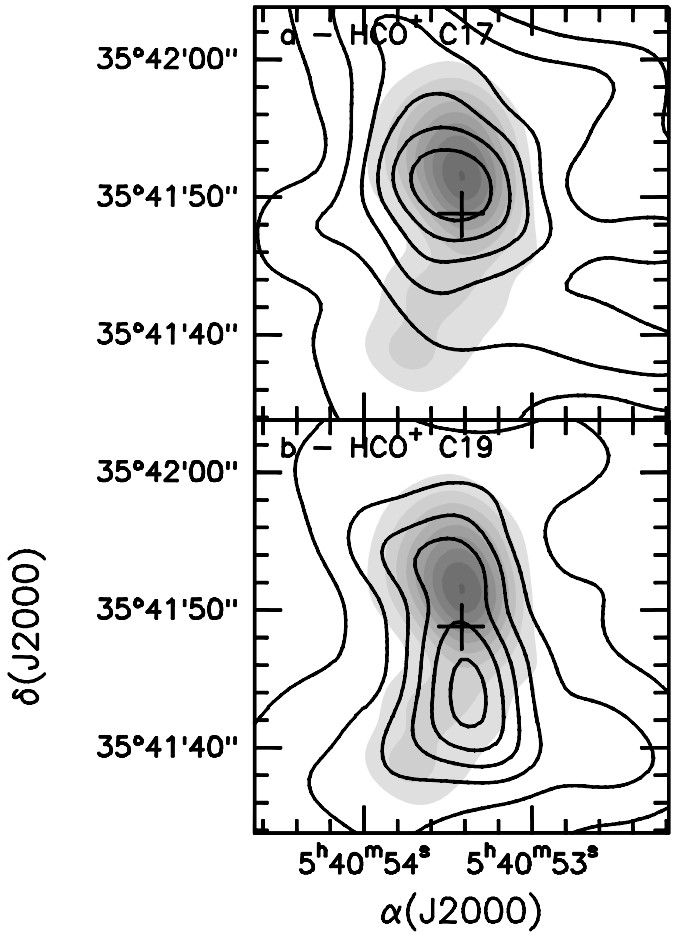,width=8.8cm}} \caption[]{Overlay of the
HCO$^+$ C17 (top) and C19 (bottom) cores (contours) with the 3.3 mm map (grey
scale). The cross marks the position of the water maser.  } \label{hcojet}
\end{figure}

\begin{figure}
\centerline{\psfig{figure=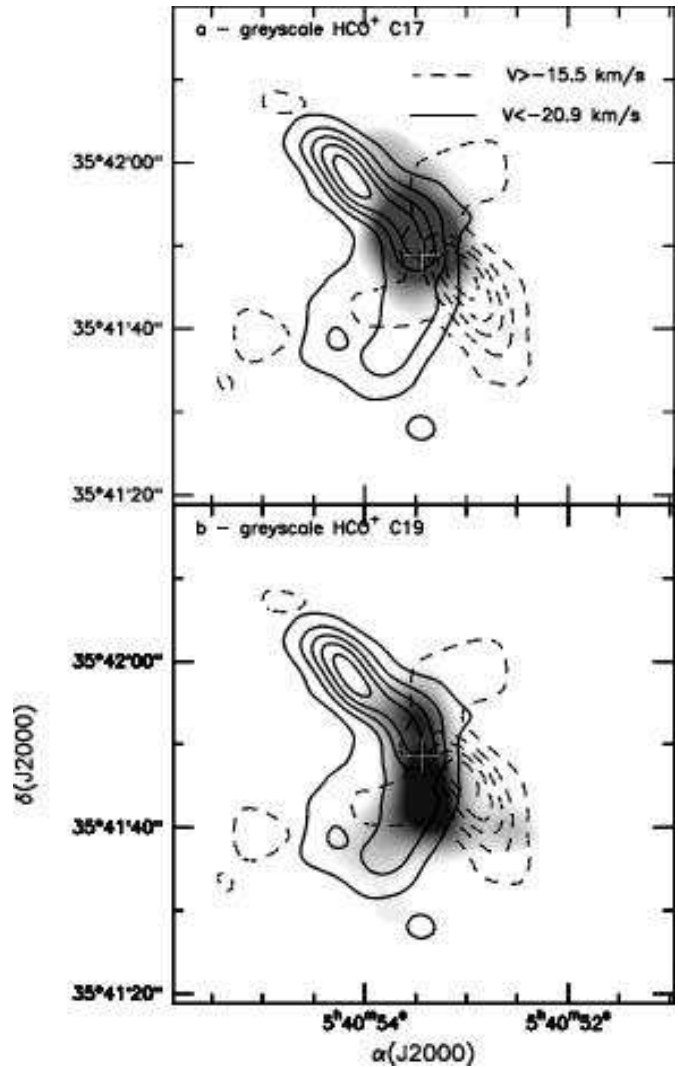,width=8.8cm}}
\caption[]{Overlay of the HCO$^+$ outflow (contours) 
with C17 (top panel) and C19   (bottom panel)
HCO$^+$ cores (grey scale). The cross
marks the position of the water maser.
}
\label{hcooutf}
\end{figure}

To study the significance of the two velocity peaks
displayed in the HCO$^+$ and CH$_3$CN spectra  and 
outlined in Sects~\ref{hco+} and~\ref{ch3cn},
we have overlaid the 
HCO$^+$ C17 and C19 cores with the 3.3 mm continuum 
emission  (Fig.~\ref{hcojet}) 
and the HCO$^+$ outflow (Fig.~\ref{hcooutf}). 

C17 is associated with the 3.3 mm core, the water maser,  and the
centre of the  NE--SW outflow. There is little doubt that the NE--SW outflow
originates from the YSO hidden within the 3.3 mm core.

C19 instead does not have an associated
3.3 mm core  and 
is located along the axis of the thermal jet
observed at 3.3 mm and the blue lobe of the  NNW--SSE outflow.
At the same position and in the same velocity interval (from
--18 to --20 \kms) one sees the jet observed in the H$_2$CS line.

The interpretation of this complex morphology is far from clear.
The most obvious one is that the molecular core is composed  
of two distinct components with different velocities.
However, in this case there would be a major difference
between C17 and C19 
due to the lack of mm continuum emission from C19: either the 
temperature of the dust in C19 is much lower than in C17 or the 
abundance of HCO$^+$ and CH$_3$CN in C19 is much higher than in C17. 
If this were  the case it would be of interest to know if the 
thermal jet and the NNW--SSE outflow originate from  C19, 
but this cannot be derived from the present data. 

An alternative explanation might  be 
that the  acceleration of the molecular gas from its interaction between the 
thermal jet and the molecular outflow 
is capable to alter in this area the velocity observed in the 
HCO$^+$ and CH$_3$CN lines, which would be the
weighted mean of the gas entrained 
and the quiescent surrounding molecular gas. In IRAS\,18162--2048 
Benedettini et al.~(\cite{bmtn}) have found that the momentum deposited
by the outflow can break the molecular cloud  apart, shifting part
of the emission towards bluer or redder velocities. 
If all the 
momentum in the blue lobe of the NNW--SSE outflow (13 \msol \kms) 
is deposited in the molecular cloud,  given the velocity offset between 
C17 and C19 (2 \kms), up to 6 \msol\ can be moved. This mass is not too
different from that of C19 (see Table~\ref{massethin}) and, if accelerated,
would definitely create a secondary velocity peak in the molecular line
emission.

%
%

\subsection{The thermal  jet}

The detection of thermal  jets in   YSOs is by no  means a new discovery
since many have been found in low luminosity YSOs 
(Anglada et al.~\cite{anglada}). Usually they are associated with molecular
outflows, even though on a much smaller scale, 
and are elongated along the axis of
the outflow. The origin of the ionization of the thermal 
jet is not photoionization,
because at the low bolometric luminosities the implied stellar 
temperatures are too low. The alternative mechanism proposed is ionization
from UV photons produced in the shock of the neutral
wind from the YSO against the surrounding  high density gas (Curiel et 
al.~\cite{curiel1},~\cite{curiel2}; Torrelles et a.~\cite{torr}).
Anglada et al.~(\cite{anglada}) have found an empirical relation between
the momentum rate in the molecular outflow ($\dot{P}$) and the radio continuum
luminosity of the jet  at centimeter wavelengths ($S_{\nu}d^2$).
Using their Eq.~(3) and the extrapolated centimeter
flux density of the thermal jet, we  derive $\dot{P}=9~10^{-3}~\msol~\kms$.
This value can be compared to that derived for the blue
lobe of the NNW--SSE outflow  using the values of Table~\ref{toutflows}, i.e.
$\dot{P}=10^{-3}~\msol \kms$. Although the two values differ by almost a factor
10, we believe that they are in reasonable agreement within the
uncertainties. In fact, on the one hand the relation by Anglada et
al.~(\cite{anglada}) has a spread of about an order of magnitude, on the other
the masses (and hence the momenta) in Table~\ref{toutflows} depend on the
HCO$^+$ abundance, which may differ significantly from the value adopted
by us.

\subsection{The disk}
\label{sdisk}

Highly collimated bipolar outflows 
have been found to be associated with disks  in several YSOs,
as for instance in HH~30 (Burrows et al. \cite{burr}) and
IRAS\,201026+4140 (Cesaroni et al~\cite{cesa1}, \cite{cesa2}).
The detected disks rotate around the central
(proto)star and lie in a plane perpendicular to the bipolar outflows.
One way to detect the disk is by  revealing the change of the 
LSR velocity along a direction perpendicular to the bipolar outflow.
To search for the possible presence of a disk in the centre of the
HCO$^+$ outflow we have used the highest resolution available
in our data and the strongest molecular line, i.e. the C$^{34}$S
observations.  

\begin{figure}
\centerline{\psfig{figure=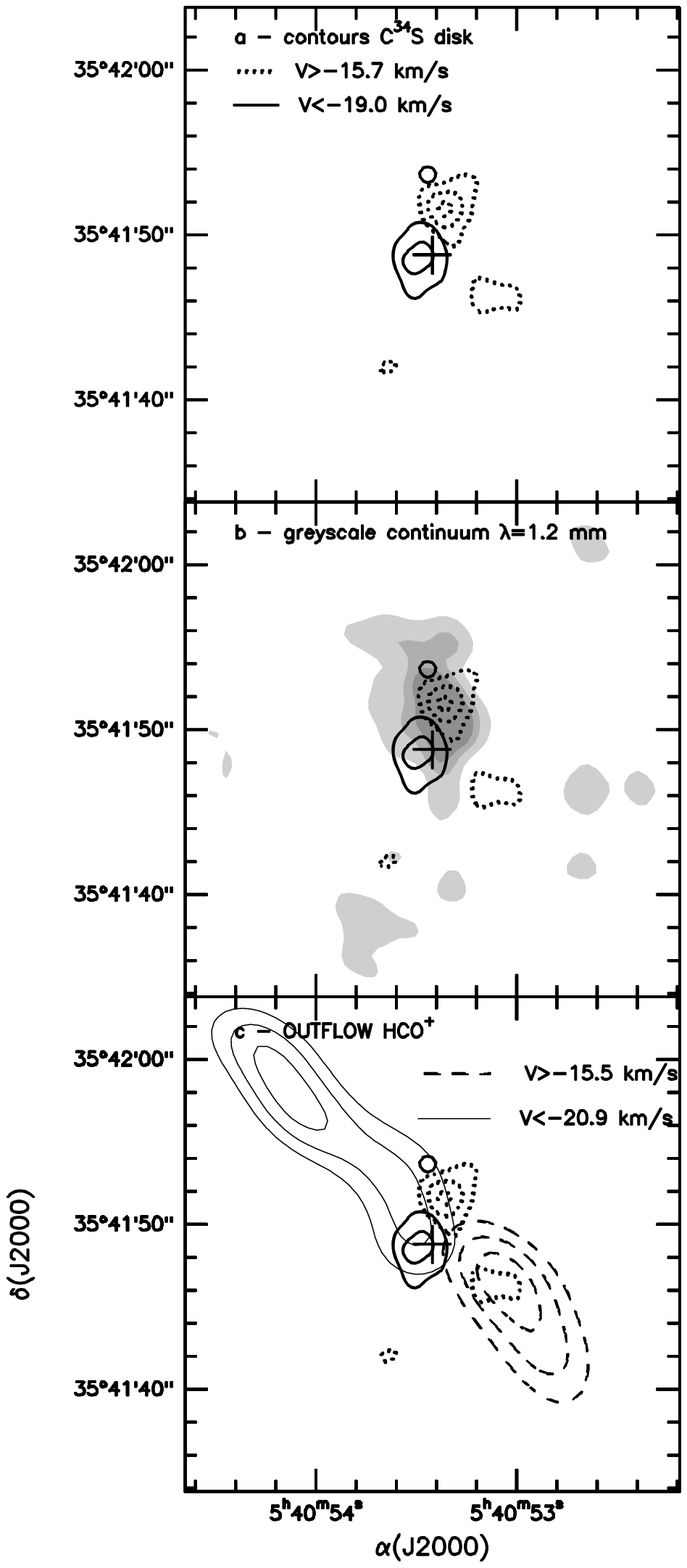,width=8.8cm}}
\caption[]{{\bf a.} Blue and red shifted emission  
of the C$^{34}$S line (contour
levels from 60 to 180 by 60 mJy/beam).
{\bf b.} Same as {\bf a} overlaid with the 1.2 mm map (grey scale).
{\bf c.} Same as {\bf a}  overlaid with the most intense part of the
HCO$^+$ outflow (contours).
The cross marks the position of the water maser.
}
\label{disk}
\end{figure}

\begin{figure}
\centerline{\psfig{figure=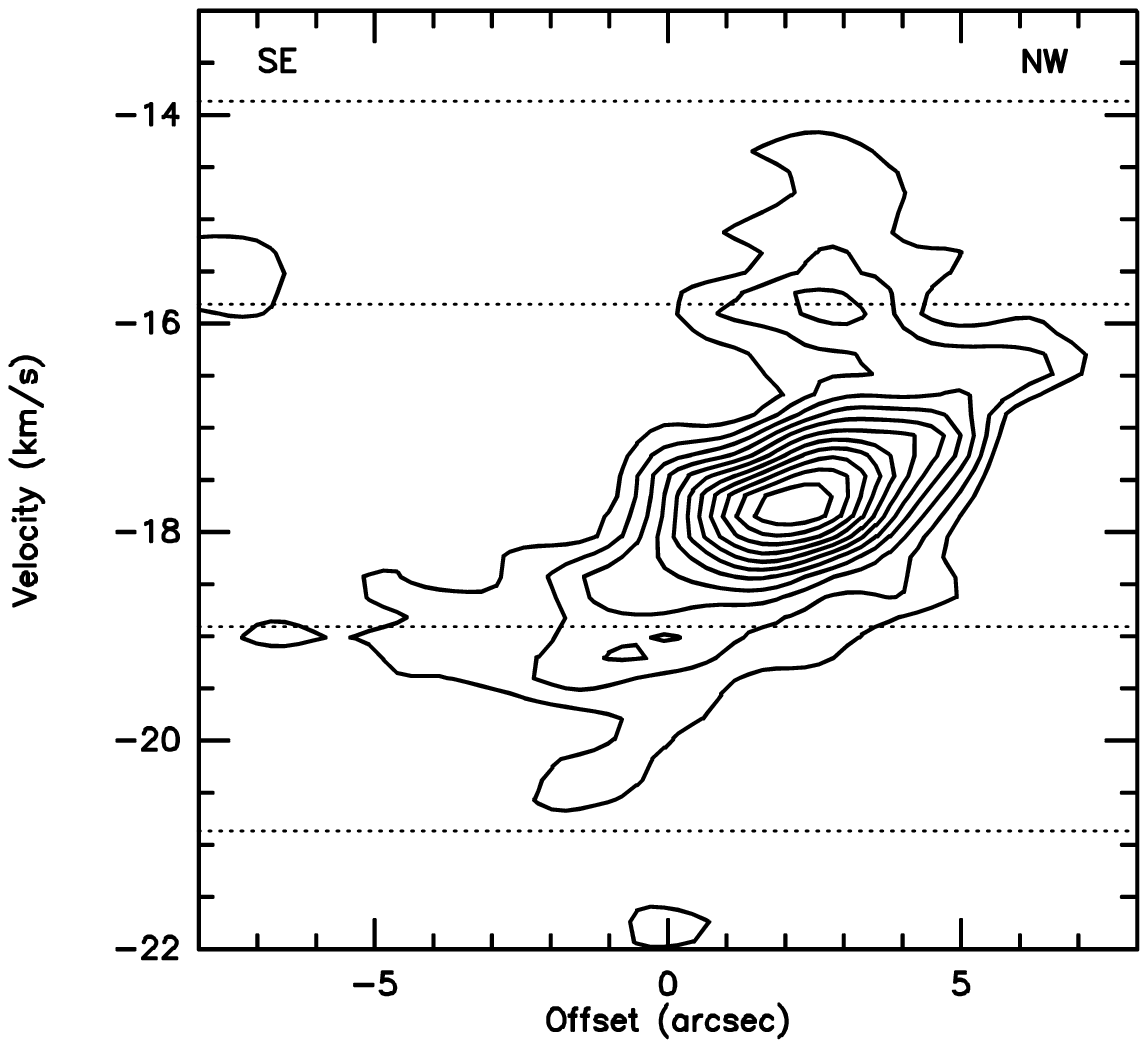,width=8.8cm,angle=-90}}
\caption[]{Position-velocity plot of the C$^{34}$S along the 
direction of the plane of the disk, averaged over a width of 4\arcsec\
in the perpendicular direction. The dotted lines define the
velocity interval which has been used to produce the maps of
Fig.~\ref{disk}. The zero offset position corresponds to the water maser.
Contours are from 0.05 to 0.55 in steps of 0.05 Jy/beam.
}
\label{vpdisk}
\end{figure}

As pointed out in Sect.~\ref{mmlines} from the comparison of single dish 
and interferometric observations, the C$^{34}$S emission is dominated by
the quiescent gas of the extended molecular cloud. 
Therefore, in order to avoid confusion
we have selected
the extreme velocity ranges where the emission is still above the
noise level, thus averaging the emission from  --20.86 to --18.9~\kms\
(blue side)  
and from --15.81 to --13.86~\kms\ (red side).  
The flux density contained
in these intervals is $\sim$17\%  of that within the 
above velocity limits.
The resulting maps are  shown in Fig.~\ref{disk}.
Two unresolved components are found (Fig.~\ref{disk}a) 
very close to the water maser and to the 1.2 mm core  (Fig.~\ref{disk}b),
which  also coincide with the centre of the NE--SW
outflow (Fig.~\ref{disk}c). 
The red and 
blue peaks are offset by $\sim$3.5\asec\ and the line that connects them
is almost perpendicular to that of the  NE--SW outflow.
This  might  suggest that one is observing a disk rotating around the YSO
and perpendicular to the outflow axis.
One could possibly object that the two unresolved components
represent the feet of the NNW--SSE outflow, given also the
good correspondence in velocity and position. While we cannot exclude this 
hypothesis, we do not feel this is the case
because the C$^{34}$S molecule is  a good tracer  
of high-density gas. To confirm our hypothesis we have made a velocity--position
plot along the direction of the plane of the disk. This is
shown in Fig.~\ref{vpdisk} which clearly indicates the existence
of a velocity gradient.

Assuming that we are dealing with a rotating disk, 
it is possible to estimate
the mass of the star+disk system from the measured radius and corresponding
rotation velocity  and check if this value is consistent
with the other indications.  
The radius  can be computed from the separation between
the red- and blue-shifted C$^{34}$S gas in Fig.~\ref{disk}c, which
at the assumed distance of 1.8~kpc corresponds to 3150~AU.
The rotation velocity is taken equal to the maximum
velocity offset between the most blue- and red-shifted
channels in which the C$^{34}$S emission is still above the noise,
and turns out to be $\sim$5~\kms. The inferred dynamical mass of the disk+star
system is $\geq$22~\msol. Although this is not a demonstration
of the existence of a disk, such a value compares well to the disk mass 
found in other YSOs  with similar mass, as for instance
IRAS\,20126+4104 (Cesaroni et al.~\cite{cesa1}).

The lower limit comes from
the unknown inclination angle of the disk with respect to
the line of sight.
This value is larger than  the disk mass derived from
the fit to  the mm flux densities. Given all the uncertainties,  
this result is consistent with the hypothesis of a disk
of mass $\sim$16 \msol, as derived from the fit to the SED
(see Sect.~\ref{score}),  rotating around a YSO of $\geq$6~\msol.

%
%

\section{The nature of the YSOs in the molecular  core}
\label{natu}

%
%

The outflows observed in the HCO$^+$ line witness the existence of embedded
YSOs powering them. The question we want to address here is whether these are
low- or high-mass stars. The molecular  
core is significantly colder (30~K) than a
``hot core'', whose temperature is typically above 100~K, and also
the line width is smaller than that found in hot cores: since hot cores are
believed to host newly born massive stars (see Kurtz et al. \cite{kcchw}),
such a temperature  and line width difference might suggest 
that one is observing lower mass
YSOs in the molecular  core. 
To find out whether this is the case, one has
to obtain a better  constrained  estimate of the luminosity of the YSOs.

The presence of two outflows guarantees that at least two distinct YSOs must
be present in the core. The possibility that multiple outflows
may originate from the same YSO, although it cannot be excluded a priori, 
seems less supported by the orthogonal position of the two 
outflows. 

According to Churchwell (\cite{churc}), the
luminosity of the star powering the outflow can be inferred from the mass
loss rate. Using his Eq.~(1) and our results from Table~\ref{toutflows}, one
obtains 3400~$L_\odot$ and 170~$L_\odot$ for the YSOs powering
respectively the NE--SW and NNW--SSE outflows. Clearly, the second cannot
contribute significantly to the total luminosity estimated in
Sect.~\ref{score}, which is dominated by the most massive YSO.

Another luminosity estimate can be calculated from the temperature obtained
from the CH$_3$CN lines. We measure 30~K at a distance of 0.03~pc (the radius
of the CH$_3$CN emitting region; see Table~\ref{massevir}) from the core centre.
If the gas is heated by a central star, then it is possible to compute the
stellar luminosity. The result depends on the geometry of the core. If this
is spherically symmetric, one can use the relation of Scoville \& Kwan
(\cite{sk}), which gives a luminosity of 220~$L_\odot$ for  
$\kappa_{\nu} \propto \nu^{0.5}$.
Much higher luminosities are predicted, up to 2700~$L_\odot$, 
if instead we use the more canonical value of  
$\kappa_{\nu} \propto \nu$ and Eq.~(31) of Goldreich \& Kwan (\cite{gokw}).
For a disk-like
geometry, instead, even an O3 star cannot justify a temperature as large as
30~K at a distance of 0.03~pc from the star.  It must be noted that this
result does not rule out the possibility that one is indeed observing a
circumstellar disk, as discussed in Sect.~\ref{sdisk}, but this must be
``active'', namely heated internally by viscosity.

Finally, one can infer the luminosity from the mass of the star estimated
in Sect.~\ref{sdisk}. The latter is $\ge$6~$M_\odot$, which implies
$\ge$1000~$L_\odot$ for a zero-age main-sequence star or alternatively
$\ge$3100~$L_\odot$ for a massive protostar (Behrend \& Maeder \cite{bema}).

All previous luminosity estimates must be compared to the value obtained from
the SED in Fig.~\ref{sed}, which ranges from 160 to 900~$L_\odot$.
However, the uncertainty in this value is not
negligible, as the SED is known only in the millimeter and sub-millimeter
regions of the spectrum. On the other hand, the other
luminosity estimates are
also affected by significant errors. In fact, the relationship used to
derive the luminosity from the mass loss rate has a spread of a factor
$\sim$3, while the value obtained from the core temperature is proportional
to $T^5$, which makes it very sensitive to small errors on $T$. In
conclusion, we believe that the estimates obtained are consistent with each
other within the uncertainties and that the YSO powering the NE--SW flow is
likely to be a B (proto)star with luminosity in the range 1000--3500~$L_\odot$,
while the NNW--SSE outflow originates from a less massive (proto)star.

We note that for this range of luminosities,  the mass infall rate derived from
the relation of  Churchwell (\cite{churc}) and  Behrend \& Maeder (\cite{bema})
turns out to be of the order of 10$^{-3}$~$M_{\odot}\,$yr$^{-1}$, not too 
different from the mass infall rate derived from the infall velocity. Even
though the
close correspondence between the two numbers should  be taken 
with caution, given the large uncertainties in the
derived quantities and the approximations used, it suggests
that gravitational energy might be the main source of luminosity for
the YSO powering the water maser.

\section{Conclusions}

The detection of a mm and sub-mm core  at the position of the water maser
in S235A-B,
as well as of a core  in all the molecular lines  finally settles
the question which is the source that provides the energy for the
maser emission and indicates that a new YSO (or a cluster of YSOs) 
is located in this region and is the powering source of the maser emission.
The two close-by nebulosities S235A-B 
do not seem to be directly related to the YSO, even though belonging to the 
same star forming complex. Consequently, star formation has not been 
occurring simultaneously 
in the extended CO molecular cloud, but rather progressing
slowly from the outer parts  of the molecular cloud, where S235A and B 
are placed, to the denser inner part of it, where new YSOs are just 
being formed.

The present example and a look at the literature on this star forming region 
indicated the complexity in finding new intermediate luminosity 
YSOs when they are
located in regions where high-mass star formation is already in progress 
and where their emission is completely obliterated by that of 
the more evolved regions, especially when the search occurs 
in the classical domains
which were originally used to find them, e.g. cm
radio continuum, near IR and  H$\alpha$ emission. 
Only water masers, as well as other masers, e.g.  
methanol (Walsh et al.~\cite{wmabl}),
have remained as good indicators of YSOs (when they are present) and 
only high resolution mm continuum and molecular line observations
are able to pinpoint the YSO emission from that of other close-by
more evolved sources (Jenness et al.~\cite{jsp}).

The detection of two bipolar outflows in S235  is another good indicator of
recent star formation. It indicates that more than one YSO might
be present or, but more difficult to explain with the present data,  
multiple episodes of bipolar outflow might
originate from the same YSO. Larger scale bipolar outflows detected
in this region 
with lower resolution might be remnants of previous episodes
of star formation, but do not seem to be directly related to the
two outflows produced by the presently detected YSOs. 
 
Sulphur-bearing molecules (H$_{2}$CS and SO$_{2}$) have been detected,
similarly to hot cores.
Part of the emission appears related with
the outflows, suggesting a shock origin, but most of it clearly arises
from the molecular core.

The luminosity of the YSO powering the NE--SW flow 
is estimated to be in the range 1000--3500~$L_\odot$,
less than that of S235A (1.1 10$^4$~\lsol) and the dynamical time 
of the order of 10$^4$~yr, much younger than that of the \HII\ region
around S235A (the canonical value for well developed \HII\ region is 10$^6$~yr).
The temperature of the molecular gas of the core
($\sim$30 K) is smaller than that
found in hot cores ($\geq$100 K) further confirming that the newly
detected YSO is intermediate between the massive objects
found in hot cores (Kurtz et al.~\cite{kcchw}) 
and the low mass class 0 objects 
(Andr\'e et al.~\cite{awb}).

In conclusion, the newly found YSO in between S235A and S235B  
is a rare example of an intermediate
luminosity object  and its morphology may be used to connect  
early evolutionary phases of massive stars with those of 
low mass protostars of class 0--I.

We find weak evidence for a possible 
disk in the C$^{34}$S line,  perpendicular to the
direction of the main bipolar outflow. 
If confirmed, this  is a further proof of how the
complex morphology of star forming regions in the earliest  phases
can be unveiled only by high resolution molecular observations in high
density tracers that can separate the weak emission of the disk
from the overwhelming emission of the molecular cloud.

While revealing a new wealth of information, 
at the same time the latest  observations leave some unanswered questions,
as for instance the origin of the thermal jet and the nature of the
two-velocity structure of the molecular cloud. 
Are there multiple YSOs in the 
region of the mm core  or a single YSO with multiple episodes
of outflows? Why there is no thermal 
jet associated with the  main bipolar
outflow? Why is the velocity of the water maser features so different
from that of the molecular core?
Work is in progress to try to give an answer to these questions. 

\begin{acknowledgements}

This research made use of data products from the Midcourse Space Experiment
(MSX).
Processing of the data was funded by the Ballistic Missile Defense Organization
with additional support from NASA Office of Space Science. IRAS HIRES images
were obtained from  the NASA/IPAC Infrared Science Archive, 
which is operated
by the Jet Propulsion Laboratory, California Institute of Technology, under
contract with the National Aeronautics and Space Administration.
The James Clerk Maxwell Telescope is operated by the Joint Astronomy
Centre in Hilo, Hawaii on behalf of the parent organizations PPARC in
the United Kingdom, the National Research Council of Canada and The
Netherlands Organization for Scientific Research. T.J. acknowledges
the support software provided by the Starlink Project which is run by
CCLRC on behalf of PPARC.

\end{acknowledgements}

\clearpage

\end{document}